\documentclass[twocolumn,english,showpacs,preprintnumbers,amsmath,amssymb,floatfix]{revtex4-1}

\usepackage[T1]{fontenc}
\usepackage[latin9]{inputenc}
\usepackage{color}
\usepackage{array}
\usepackage{amstext}
\usepackage{graphicx}
\usepackage{esint}
\usepackage{rotating}
\usepackage{appendix}
\usepackage{float}
\usepackage{ulem}

\usepackage[font=small,labelfont=bf]{caption}
\usepackage{xcolor}
\usepackage[colorlinks = true,
linkcolor = blue,
urlcolor  = blue,
citecolor = blue,
anchorcolor = blue]{hyperref}

\makeatletter

\providecommand{\tabularnewline}{\\}

\@ifundefined{textcolor}{}
{%
 \definecolor{BLACK}{gray}{0}
 \definecolor{WHITE}{gray}{1}
 \definecolor{RED}{rgb}{1,0,0}
 \definecolor{GREEN}{rgb}{0,1,0}
 \definecolor{BLUE}{rgb}{0,0,1}
 \definecolor{CYAN}{cmyk}{1,0,0,0}
 \definecolor{MAGENTA}{cmyk}{0,1,0,0}
 \definecolor{YELLOW}{cmyk}{0,0,1,0}
 }

\@ifundefined{definecolor}
 {\usepackage{color}}{}
\@ifundefined{definecolor}
 {\usepackage{color}}{}
\makeatother
\usepackage{babel}

\begin{document}




\title { Analytical approaches to the determination of spin-dependent parton distribution functions at NNLO approximation }


\author{Maral Salajegheh$^{1}$}
\email{M.Salajegheh@stu.yazd.ac.ir}

\author{S. Mohammad Moosavi Nejad$^{1,3}$}
\email{mmoosavi@yazd.ac.ir}

\author{Hamzeh Khanpour$^{2,3}$}
\email{Hamzeh.Khanpour@mail.ipm.ir}

\author{S. Atashbar Tehrani$^{4}$}
\email{Atashbart@gmail.com}

\affiliation {
$^{(1)}$Physics Department, Yazd University, P.O. Box 89195-741, Yazd, Iran      \\
$^{(2)}$Department of Physics, University of Science and Technology of Mazandaran, P.O.Box 48518-78195, Behshahr, Iran    \\
$^{(3)}$School of Particles and Accelerators, Institute for Research in Fundamental Sciences (IPM), P.O.Box 19395-5531, Tehran, Iran   \\
$^{(4)}$Independent researcher, P.O. Box 1149-834413, Tehran, Iran   }

\date{\today}

%
%
\begin{abstract}\label{abstract}

In this paper, we present ``{\tt SMKA18}'' analysis which is a first attempt to extract the set of next-to-next-leading-order (NNLO) spin-dependent parton distribution functions (spin-dependent PDFs) and their uncertainties determined through the Laplace transform technique and Jacobi polynomial approach.
Using the Laplace transformations, we present an analytical solution for the spin-dependent Dokshitzer-Gribov-Lipatov-Altarelli-Parisi evolution equations at NNLO approximation. The results are extracted using a wide range of proton $g_1^{p}(x, Q^2)$, neutron $g_1^{n}(x, Q^2)$ and deuteron $g_1^{d}(x, Q^2)$ spin-dependent structure functions dataset including the most recent high-precision measurements from {\tt COMPASS16} experiments at CERN which are playing an increasingly important role in global spin-dependent fits. The careful estimations of uncertainties have been done using the standard ``Hessian error'' propagation. We will compare our results with the available spin-dependent inclusive deep inelastic scattering dataset and other results for the spin-dependent PDFs in literature. The  results obtained for the spin-dependent PDFs as well as spin-dependent structure functions are clearly explained both in the small and large values of $x$.

\end{abstract}

\maketitle

%
%
\section{Introduction}\label{Introduction}

The structure of hadrons, specifically unpolarized parton distribution functions PDFs~\cite{Ball:2017nwa,Bourrely:2015kla,Harland-Lang:2014zoa,Hou:2017khm,Alekhin:2017kpj,Eskola:2016oht,Kovarik:2015cma,Selyugin:2014sca,Selyugin:2015pha,Goharipour:2017uic,Goharipour:2017rjl,Rostami:2016dqi,Aleedaneshvar:2016dzb,Aleedaneshvar:2016rkm} and spin-dependent PDFs dynamics~\cite{Shahri:2016uzl,Khanpour:2017fey,Khanpour:2017cha,Jimenez-Delgado:2014xza,Nocera:2014gqa,Sato:2016tuz,Jimenez-Delgado:2013boa}, is an interesting topic which continues to attract more attention from large physics communities~\cite{Goto:1999by}.

From the practical point of view, experiments including nucleon beams at the current and future energy frontiers need most accurate information on the spin-dependent PDFs
to exploit their data. In the absence of improved inclusive deep inelastic scattering (DIS) data, most  attention is now turned towards data from the Large Hadron Collider (LHC) experiments.
In addition, the nucleon spin structure has been always a fundamental question in high energy physics so it has been extensively studied both in theory and experiment in recent decades.
From an experimental point of view, several experiments have been set up to study the longitudinal spin structure of the nucleon, such as COMPASS experiments at CERN~\cite{Alekseev:2010hc,Adolph:2015saz,Adolph:2016myg}, {\tt HERMES} experiments at DESY~\cite{Airapetian:2006vy,LopezRuiz:2012hxr,Schnell:2013sla}, many experiments at Jefferson Laboratory ({\tt JLAB})~\cite{Kramer:2005qe,Guler:2015hsw,Flay:2016wie}, and {\tt PHENIX} and {\tt STAR} experiments at the proton-proton ($pp$) Relativistic Heavy Ion Collider ({\tt RHIC})~\cite{Adamczyk:2016okk,Adare:2016cqe,Adare:2015ozj,Adare:2015gsd}. The majority of this experimental information on spin-dependent PDFs come from the neutral-current inclusive and semi-inclusive deep inelastic scattering, DIS and SIDIS, with charged lepton beams $\ell^\pm$ and nuclear targets. These inclusive spin-dependent DIS data constrain only the total quark combinations, while the SIDIS data constrain individual quark and anti-quark flavors in the nucleon. In principle, both DIS and SIDIS data are also sensitive to the gluon distribution, however, the constraining power of DIS and SIDIS data on the gluon distribution is rather weak due to the limited kinematical range covered by these data sets.

Using the available and up-to-date datasets, many various global QCD analyses of nucleon spin structure at next-to-leading (NLO) accuracy have been  completed, recently.
These spin-dependent PDF analyses incorporate, NNPDF Collaboration {\tt NNDPFpol1.1}~\cite{Nocera:2014gqa}, {\tt DSSV09}~\cite{deFlorian:2009vb}, Asymmetry Analysis Collaboration {\tt AAC09}~\cite{Hirai:2008aj}, {\tt BB10}~\cite{Blumlein:2010rn}, {\tt LSS10}~\cite{Leader:2010rb}, {\tt DSSV14}~\cite{deFlorian:2014yva} and the recent analyses from Jefferson Lab (JAM) Collaboration, ({\tt JAM13}~\cite{Jimenez-Delgado:2013boa}, {\tt JAM14}~\cite{Jimenez-Delgado:2014xza},  {\tt JAM15}~\cite{Sato:2016tuz}) and the most recent next-to-next-leading-order (NNLO) QCD analyses from {\tt TKAA16}~\cite{Shahri:2016uzl}, {\tt KTA-I-17}~\cite{Khanpour:2017cha} and {\tt KTA-II-17}~\cite{Khanpour:2017fey}.
These NLO and NNLO spin-dependent PDF analyses are based on either the spin-dependent inclusive DIS, or combined DIS and SIDIS data, or the spin-dependent proton-proton $pp$ scattering at {\tt RHIC}. These efforts show the specific challenge of global QCD analyses to incorporate a large volume of data from many experiments.

In our previous study, {\tt KTA-I-17}~\cite{Khanpour:2017cha}, we performed a first analysis of spin-dependent inclusive DIS by taking into account the contributions from target mass corrections (TMCs) and higher twist terms (HT) to the spin-dependent $g_1$ and $g_2$ structure functions. In {\tt KTA-II-17}~\cite{Khanpour:2017fey} we considered the effects of nuclear corrections such as Fermi motions, spin depolarizations, binding and the presence of
a non-nucleonic degree of freedom, shadowing and anti-shadowing corrections which are necessary at these kinematics. It is worth mentioning here that, many spin-dependent PDFs analyses impose more stringent cuts on the photon virtuality ${\text Q}^2$ as well as ${\text W}^2$ in order to avoid dealing  with the complications associated with the HTs and nuclear corrections. Unfortunately, these restrictions eliminate much of data at the highest $x$-values.
Most of these phenomenological spin-dependent PDFs analyses also utilize standard PDFs fitting technology in which single fits are performed by assuming the basic functional forms for the PDFs. In this approach, the fit parameters are obtained by minimizing the overall $\chi^2_{\rm global}$. Then, the PDFs errors are  typically computed using the Hessian method or Lagrange multiplier or neural network. In the present paper, we construct for the first time a set of spin-dependent PDFs at NNLO approximation using a methodology, the so-called ``Laplace transform technique'' and ``Jacobi polynomials approach'', which has been recently used to study the polarized~\cite{Shahri:2016uzl} and unpolarized PDFs~\cite{Khanpour:2016uxh,MoosaviNejad:2016ebo}.

As in {\tt KTA-I-17}~\cite{Khanpour:2017cha} and {\tt KTA-II-17}~\cite{Khanpour:2017fey} analyses, we use the spin-dependent inclusive DIS data whenever available. We include new data sets with high-precision {\tt COMPASS}~\cite{Adolph:2015saz,Adolph:2016myg} measurements at CERN on the proton and deuteron. To isolate the impact of recent {\tt COMPASS} data sets more directly, and to assess the systematics of our new methodology based on Laplace transform technique and Jacobi polynomials approach, we restrict the current analysis to the inclusive DIS data. A full global QCD analysis of all available data including SIDIS, inclusive jet and weak boson production in polarized proton-proton collisions, will be presented in our forthcoming studies.

The structure of our paper is as follows:
An analytical solution for the NNLO spin-dependent Dokshitzer-Gribov-Lipatov-Altarelli-Parisi(DGLAP)evolution equations is presented in Sec.~\ref{DGLAP-equations}. This section also includes the nonsinglet, singlet and gluon solutions in the Laplace space at the NNLO approximation.
The theoretical calculations for the leading-twist spin-dependent DIS structure function and Jacobi polynomials approach used in the {\tt SMKA18} spin-dependent PDFs fit are summarized in Section~\ref{structure-function}. The dataset used to determine the {\tt SMKA18} spin-dependent PDFs are briefly summarized in Sec.~\ref{data}.
Section~\ref{pdf}, includes the choice of input scale and the {\tt SMKA18} parametrization basis. The results of the present spin-dependent PDFs analysis are given in Sec.~\ref{result}. This section includes a detailed comparison between the present results and available spin-dependent inclusive DIS data. We also present a detailed comparison of our NNLO results with recent results in literature. Finally Sec.~\ref{Summary} includes our summary and conclusions.
In Appendix~\ref{AppendixA}, we present the analytical solutions for the NNLO splitting functions in Laplace space. Appendix~\ref{AppendixB} includes the coefficient functions of singlet distributions in the Laplace space.

%
%
\section{ Analytical solution for the helicity-dependent DGLAP evolution equations } \label{DGLAP-equations}

We perform our analysis in Laplace $s$-space which has the advantage of significantly shorter fitting time compared to the $x$- or Mellin-space based analyses.
Following this, we describe the novel aspect of our analysis, namely the Laplace technique. It turns out, however, that in this method the computational time can be significantly reduced through the use of Laplace $s$-space techniques. This is due to that, firstly, the ${\text Q}^2$ evolution equations in Laplace space are the ordinary coupled differential equations which are faster and also simpler to be solved in comparison to the corresponding integro-differential equations in the $x$-space. Secondly, using this technique it is possible to cast the various multidimensional integrations in terms of precomputed quantities, which can significantly decrease the computational time needed for the observables in the global QCD fits.

Here, we follow the Laplace technique described in our previous unpolarized PDFs analyses~\cite{MoosaviNejad:2016ebo,Khanpour:2016uxh} to present an analytical solution for the coupled NNLO DGLAP evolution equations.
This method has been successfully developed and used in variety of QCD analyses, (see Refs.~\cite{MoosaviNejad:2016ebo,Khanpour:2016uxh,Block:2010du,Block:2010fk,Block:2009en,Block:2010ti,Block:2007pg,Block:2008xc,Zarrin:2016kxf,Boroun:2016uan,Boroun:2016zql,Boroun:2015hiy,Boroun:2015cta,Boroun:2015aya,Boroun:2014zpa,Boroun:2013mgv,Boroun:2014dka,Mottaghizadeh:2016krr,Taghavi-Shahri:2016ktz,Devee:2012zz,Shah:2008zz,Baishya:2009zz,Baishya:1900zz,Zarei:2015jvh,Pari:2015gta,AtashbarTehrani:2013qea,Block:2011xb} for clear reviews). We will show that this method can be also used in the spin-dependent case and, hence, one can extract the spin-dependent PDFs inside the nucleon from QCD analysis of spin-dependent inclusive DIS data. Here,  we will give the details of our approach and review the method of extracting the spin-dependent PDFs. It must be noted that, first, we will focus on the nonsinglet solution at NNLO approximation in Laplace space and, second, we will present our analytical solution for the singlet and gluon cases.

%
%
\subsection{ Nonsinglet solution in Laplace space at the NNLO approximation } \label{NonsingletLaplace}

According to the decoupling of the DGLAP evolution equations in the Laplace method, used in {\tt SMKA18} QCD analysis, we convert the solutions into three parts which take the nonsinglet $\Delta f_{\rm NS}$, singlet quark $\Delta f_{\rm S}$ and gluon $\Delta f_{g}$ distributions. In this section, we present our solution for the nonsinglet sector and left the singlet quark and gluon ones for the next section.

As was mentioned in the ~\ref{Introduction}, the DGLAP evolution equations~\cite{Dokshitzer:1977sg,Gribov:1972ri,Lipatov:1974qm,Altarelli:1977zs} are a set of integro-differential equations which can evolve the PDFs into a desired ${\text Q}^2$-scale. The nonsinglet sector of the DGLAP evolution equations at NNLO approximation reads:
\begin{eqnarray}
\label{eq:nonsinglet-DGLAP}
&&\frac{4 \pi}{\alpha_s(Q^2)}
\frac{\partial\Delta q_{\text{NS}}}
{\partial \ln Q^2} (x,Q^2) =
\Delta q_{\text{NS}} \otimes
\Big(\Delta p_{\text{NS}}^{\text{LO}}
+ \nonumber  \\
&&\frac{\alpha_s(Q^2)}
{4\pi}\Delta p_{\text{NS}}^{\text{NLO}}
+(\frac{\alpha_s(Q^2)}{4\pi})^2
\Delta p_{\text{NS}}^{\text{NNLO}}
\Big)(x,Q^2),
\end{eqnarray}
where $\Delta q_{\text{NS}}$ stands for the nonsinglet spin-dependent PDFs, the symbol $\otimes$ denotes the convolution and $\alpha_s(Q^2)$ is the QCD coupling constant.
The parameters $\Delta p_{_{\text{NS}}}^{\text{LO}}[\alpha_s(Q^2)]$, $\Delta p_{_{\text{NS}}}^{\text{NLO}}[\alpha_s(Q^2)]$ and $\Delta p_{_{\text{NS}}}^{\text{NNLO}}[\alpha_s(Q^2)]$ are the nonsinglet spin-dependent Altarelli-Parisi splitting kernels at one, two and three loops corrections, respectively.

Let us now briefly review the method of extracting the spin-dependent PDFs using the analytical solutions of DGLAP evolution equations applying the
Laplace transformation technique. Considering the variable definitions as $\nu \equiv \ln(1/x)$ and $w \equiv \ln(1/z)$, one can rewrite the evolution equations \eqref{eq:nonsinglet-DGLAP} in terms of the convolution integrals and also the new variables $\nu$ and $w$.
Therefore, one can obtain a simple solution as follows:
\begin{eqnarray}\label{eq:nonsinglet}
&&\frac{\partial \Delta \hat{F}_{\text{ NS}}}
{\partial\tau}(\nu, \tau)
= \int_0^\nu   \Big
(\Delta p_{\text{NS}}^{\text{LO}}(\nu-w)
+ \nonumber \\
&&\frac{\alpha_s(\tau)}{4 \pi}
\Delta p_{\text{NS}}^{\text{NLO}}(\nu-w)+
(\frac{\alpha_s(\tau)}{4 \pi})^2
\Delta p_{\text{NS}}^{\text{ NNLO, NS}}
(\nu-w) \Big)  \nonumber \\
&&\Delta \hat{F}_{\text{ NS}}
(w, \tau)e^{-(\nu - w)}\,dw \,.
\end{eqnarray}
It is worth noting here that the ${\text Q}^2$-dependence of the above evolution equations is expressed thorough the variable $\tau$ as $\tau (Q^2, Q_0^2) \equiv \frac{1}{4 \pi} \int_{Q_0^2}^{Q^2} \alpha_s ({Q^{\prime}}^{2}) d\ln {Q^{\prime}}^{2}$.

By defining the Laplace transforms as $\Delta f_{\text {NS}}(s, \tau) \equiv  {\cal L} [\Delta \hat F_{\text {NS}}(\nu, \tau); s]$ and considering the fact that the Laplace transform of convolution factors is simply the ordinary product of the Laplace transform of the factors~\cite{Block:2010du,Block:2011xb}, the Laplace transform of Eq.~\eqref{eq:nonsinglet} leads to the ordinary first order differential equations in Laplace space $s$ with respect to the $\tau$-variable.
Therefore, by working in the Laplace $s$-space, one can obtain the first order differential equations for the nonsinglet distributions $\Delta f_{\text {NS}}(s, \tau)$, as
\begin{eqnarray}
\label{eq:nonsinglet-laplace-space}
\frac{\partial \Delta
f_{\text{NS}}}{\partial \tau}(s, \tau) &&
= \left (\Delta \Phi_{\text{NS}}^{\text{LO}} +
\frac{\alpha_s(\tau)} {4 \pi}
\Delta\Phi_{\rm NS}^{\text{NLO}} +
\right.\nonumber  \\
&&\left.(\frac{\alpha_s(\tau)}
{4 \pi})^2\Delta \Phi_{\text{NS}}^
{\text{NNLO}} \right)
\times\Delta f_{\text{NS}}(s, \tau)\,.
\end{eqnarray}
A very simplified solution for the above equation is
\begin{eqnarray}\label{eq:solve-nonsinglet}
\Delta f_{\text{ NS}}(s, \tau) =
e^{\tau\Delta\Phi_{\text{NS}}(s)}
\, \Delta f^0_{\text{NS}}(s) \,,
\end{eqnarray}
where, up to the NNLO approximation, the $\Phi_{\text{NS}}(s)$ contains contributions of the splitting functions at the $s$-space. Up to NNLO approximation, it reads
\begin{eqnarray}\label{eq:fi-nonsinglet}
\Delta\Phi_{\text{NS}}(s)
\equiv \Delta
\Phi_{\text{NS}}^{\text{LO}}(s) +
\frac{\tau_2}{\tau} \Delta
\Phi_{\text{NS}}^{\text{ NLO}}(s) +
\frac{\tau_3}{\tau}
\Delta\Phi_{\text{NS}}^{\text{NNLO}}(s)\,. \nonumber \\
\end{eqnarray}
These splitting functions can be calculated from $x$-space results and presented in Refs.~\cite{Lampe:1998eu,Moch:2014sna}. The ${\text Q}^2$-dependence variables $\tau_2$ and $\tau_3$ in Eq.~\eqref{eq:fi-nonsinglet} are defined as follows
\begin{equation}
\tau_2 \equiv \frac{1}{(4 \pi)^2}
\int_{Q_0^2}^{Q^2}
\alpha_s^2(Q'^2)d \ln Q'^2\,,
\end{equation}
and
\begin{equation}
\tau_3 \equiv \frac{1}{(4 \pi)^3}
\int_{Q_0^2}^{Q^2} \alpha_s^3
(Q'^2)d \ln Q'^2\,.
\end{equation}
The LO and NLO contributions to the splitting function $\Delta \Phi_{\text{NS}}^{\text{LO}}(s)$ and $\Delta \Phi_{\text{NS}}^{\text{NLO}}(s)$ are presented in Ref.~\cite{AtashbarTehrani:2013qea} and the NNLO one, i.e. $\Delta \Phi_{\text{NS}}^{\text{NNLO}}(s)$, calculated in this analysis and presented in Appendix~\ref{AppendixA}.

%
%
\subsection{Singlet and gluon solutions at the NNLO approximation}\label{singlet-case}

In this section, we turn to present our solutions for the singlet quark and gluon evolutions in Laplace $s$-space. The coupled NNLO DGLAP equations, using the convolution symbol $\otimes$, for the singlet quark $\Delta F_{s}$ can be schematically written as
\begin{widetext}
\begin{eqnarray}
\label{eq:fs-singlet}
\frac{4 \pi } {\alpha_s (Q^2)}
\frac {\partial \Delta F_{s}}{\partial \ln
Q^{2}}(x, Q^2)
&=& \Delta F_s \otimes \left(\Delta P_{qq}^0
+ \frac {\alpha_s (Q^2)}{
4 \pi } \Delta P_{qq}^1 \right.
 \left. + \left(\frac
{\alpha_s (Q^2)} {4\pi } \right)
^{2} \Delta P_{qq}^2 \right) (x, Q^2)  \nonumber \\
&& + \Delta G \otimes \left (\Delta P_{qg}^0 +
\frac {\alpha_s(Q^2)} {4 \pi }
\Delta P_{qg}^1 \right.
 \left. + \left (\frac{\alpha_s (Q^2)}
{4\pi} \right)^2 \Delta
P_{qg}^2 \right) (x, Q^2) \,.
\end{eqnarray}
The corresponding evolution for the gluon density $\Delta G$ at NNLO approximation is given by
\begin{eqnarray}\label{eq:g-singlet}
\frac{4 \pi }{\alpha_s (Q^2)}
\frac{\partial \Delta G}{\partial \ln Q^2
}(x, Q^2)
&=& \Delta F_s \otimes
\left (\Delta P_{gq}^0
+ \frac{\alpha_s (Q^2)}{
4 \pi}\Delta P_{gq}^1 \right.
 \left. + \left(\frac{\alpha_s(Q^2)}
{4 \pi} \right)
^2 \Delta P_{gq}^2 \right)
(x, Q^2)  \nonumber \\
&& + \Delta G \otimes
\left(\Delta P_{gg}^0 +
\frac{\alpha_s(Q^2)} {4 \pi}
\Delta P_{gg}^1 \right.
 \left. + \left(\frac{\alpha_s(Q^2)}
{4\pi} \right)^2 \Delta
P_{gg}^2 \right) (x, Q^2) \,.
\end{eqnarray}
In the two above equations, $\Delta P_{qq}^0$, $\Delta P_{qg}^0$, $\Delta P_{gq}^0$ and $\Delta P_{gg}^0$  are the
LO singlet splitting functions.
Correspondingly, $\Delta P_{qq}^1$, $\Delta P_{qg}^1$, $\Delta P_{gq}^1$ and $\Delta P_{gg}^1$ are the NLO splitting functions, and $\Delta P_{qq}^2$, $\Delta P_{qg}^2$, $\Delta P_{gq}^2$ and $\Delta P_{gg}^2$ are the NNLO ones. The $\alpha_s (Q^2)$ is the NNLO  strong coupling constant.

As for the nonsinglet sector presented in Sec.~\ref{NonsingletLaplace}, considering new variables $\tau \equiv \frac{1}{4 \pi} \int_{Q_0^2}^{Q^2} \alpha_s({Q^\prime}^2) d\ln {Q^\prime}^2$,  $x = e^{-v}$ and $z = e^{-w}$, the DGLAP evolutions in Eqs.~\eqref{eq:fs-singlet} and \eqref{eq:g-singlet} can be written in terms of the variables $\tau$, $v$, and $w$.
Following the Laplace transformation technique, the convolution integrals in DGLAP evolutions can be converted from $v$ space to $s$ space~\cite{Block:2010du,AtashbarTehrani:2013qea}. In this respect, one can achieve the following equations for the quark singlet $\Delta f (s, \tau)$:
\begin{eqnarray}
\label{eq:fsta-singlet}
\frac{\partial \Delta f} {\partial \tau }(s, \tau)
 &=&\left( \Delta \Phi
_{f}^{\text{LO}}(s) +
 \frac{\alpha_s(\tau )}{4 \pi }\Delta \Phi
_{f}^{\text{NLO}}(s)\right.
\left. + \left( \frac{\alpha_s (\tau )}
{4 \pi }\right)^2 \Delta \Phi
_{f}^{\text{NNLO}}(s) \right)
 \Delta f(s, \tau )  \nonumber\\
&& + \left( \Delta \Theta_f^{\text{LO}}(s)
 + \frac{\alpha_s (\tau )}{4 \pi} \Delta
\Theta_f^{\text{NLO}} (s) \right.
 \left. + \left(\frac{\alpha_s(\tau )}
{4 \pi } \right)
^2 \Delta \Theta_f^{\text{NNLO}}(s) \right)
 \Delta g(s, \tau ) \,,
\end{eqnarray}
and for the gluon $\Delta g(s, \tau)$, one has
\begin{eqnarray}
\label{eq:gta-singlet}
\frac{\partial \Delta g}
{\partial \tau }(s, \tau )
&=& \left( \Delta \Phi
_{g}^{\text{LO}}(s) +
 \frac{\alpha_s (\tau)} {4 \pi } \Delta \Phi
_{g}^{\text{NLO}}(s) \right.
 \left. + \left(\frac{\alpha_s (\tau )}{4 \pi}
 \right)^2 \Delta \Phi
_{g}^{\text{NNLO}}(s) \right) \Delta g(s, \tau )        \nonumber \\
&& + \left(\Delta \Theta_g^{\text{LO}}(s) +
 \frac{\alpha_s (\tau)}{4 \pi} \Delta
\Theta_g^{\text{NLO}}(s) \right.
 \left. + \left(\frac{\alpha_s
(\tau)}{4 \pi} \right)
^2 \Delta \Theta_g^{\text{NNLO}}(s) \right)
 \Delta f(s, \tau ) \,,
\end{eqnarray}
which are expressed in terms of new variable $\tau$.

We are now in a position to extend our calculations to the NNLO approximation for the singlet and gluon spin-dependent parton distributions. To decouple and solve the DGLAP
evolutions of Eqs.~\eqref{eq:fsta-singlet} and \eqref{eq:gta-singlet}, we need an extra Laplace transformation from $\tau$ space to $U$ space. From now on, for simplicity the quantity  $\alpha_{s}(\tau)/4\pi$ is replaced by a new variable $a(\tau)$.
Therefore, the coupled DGLAP evolution equations can be converted to the following forms for the singlet and gluon distributions
\begin{eqnarray}
\label{eq:fsU-singlet}
U\Delta \mathcal{F}(s, U) - \Delta f_s^0 (s) &=&
\Delta \Phi_f^{\text{LO}}(s) \Delta \mathcal{F}(s, U)
 + \Delta \Phi_f^{\text{NLO}} (s)
 \mathcal{L}[a (\tau ) \Delta f_s (s, \tau ); U]
 + \Delta \Phi_f^{\text{NNLO}}(s)
 \mathcal{L}[a(\tau)^2 \Delta f_s (s, \tau ); U]
\nonumber \\
&& + \Delta \Theta_f^{\text{LO}}(s)
\Delta \mathcal{G}(s, U)
 + \Delta \Theta_f^{\text{NLO}}(s)
 \mathcal{L}[a(\tau) \Delta g(s, \tau); U]
 + \Delta \Theta_f^{\text{NNLO}}(s)
 \mathcal{L}[a(\tau)^2 \Delta g(s, \tau); U]\nonumber\\
\end{eqnarray}

\begin{eqnarray}
\label{eq:gU-singlet}
U \Delta \mathcal{G}(s, U) - \Delta g^0(s) &=&
\Delta \Phi_g^{\text{LO}}(s) \Delta
\mathcal{G}(s, U)
 + \Delta \Phi_g^{\text{NLO}}(s)
\mathcal{L}[a(\tau) \Delta g(s, \tau); U]
 + \Delta \Phi_g^{\text{NNLO}}(s)
 \mathcal{L}[a(\tau)^2 \Delta g(s, \tau); U]
\nonumber  \\
&& + \Delta \Theta_g^{\text{LO}}(s)
 \Delta \mathcal{F}(s, U)
+ \Delta \Theta_g^{\text{NLO}}(s)
 \mathcal{L}[a (\tau) \Delta f_s(s, \tau); U]
+ \Delta \Theta_g^{\text{NNLO}}(s)
 \mathcal{L}[a (\tau)^2
 \Delta f_s(s, \tau); U].\nonumber \\
\end{eqnarray}

In writing the above expressions, we used the conventions introduced in Ref.~\cite{Block:2010du,MoosaviNejad:2016ebo,Khanpour:2016uxh}.
At the NNLO approximation, $a(\tau)=a_{0}+a_{1}e^{-b_{1}\tau}$ is an excellent parametrization which is accurate to $\mathcal{O} (10^{-4})$~\cite{MoosaviNejad:2016ebo,Khanpour:2016uxh}.
Using this approximation for the $a(\tau)$, one can write the Laplace transforms of $\mathcal{L}[a(\tau) \Delta f_s(s, \tau); U]$, $\mathcal{L}[a(\tau)^2 \Delta f_s(s, \tau ); U]$, $\mathcal{L}[a(\tau) \Delta g(s, \tau); U]$ and $\mathcal{L}[a(\tau)^2 \Delta g(s, \tau); U]$ as
\begin{eqnarray}\label{eq:la1}
\mathcal{L}[a(\tau) \Delta
f_s(s, \tau); U]
 &=&a_0 \Delta \mathcal{F}
(s, U) + a_1 \Delta
 \mathcal{F}(s, U + b_1)
  \,, \nonumber  \\
\mathcal{L}[a(\tau)
 \Delta g(s, \tau); U]
&=&a_0 \Delta \mathcal{G}
(s, U) + a_1  \Delta
\mathcal{G}(s, U + b_1)] \,,   \nonumber\\
\end{eqnarray}
and
\begin{eqnarray}
\label{eq:la2}
\mathcal{L}[a(\tau)^2
 \Delta f_s(s, \tau); U]
&=&a_0 ^2 \Delta \mathcal{
F}(s, U) + a_1^2 \Delta
\mathcal{F}(s, U + 2 b_1)
 + 2a_0 a_1 \Delta \mathcal{F}
(s, U + b_1) \,,  \nonumber \\
\mathcal{L} [a(\tau)^2 \Delta
 g(s, \tau); U]
&=&a_0^2 \Delta  \mathcal{G}
(s, U) + a_1^2 \Delta \mathcal{G}
(s, U + 2 b_1)
 + 2 a_0  a_1 \Delta \mathcal{G}
(s, U + b_1) \,.
\end{eqnarray}

By introducing the following simplifying notations:
\begin{eqnarray}
\label{eq:la3}
\Delta \Phi_f & \equiv &
\Delta \Phi_f ^{\text{ LO}}(s) +
 a_0 \Delta \Phi
_{f}^{\text{ NLO}}(s) +
 a_0^2 \Delta
  \Phi_f^{\text{ NNLO}}(s)  \nonumber \\
\Delta \Phi_g &\equiv &
\Delta \Phi_g^{\text{ LO}}(s) +
 a_0  \Delta \Phi
_g^{\text{ NLO}}(s) +
a_0^2 \Delta \Phi_g^{\text{ NNLO}}(s)  \nonumber \\
\Delta \Theta_f
 &\equiv &\Delta \Theta_f^{\text{ LO}}(s)
  +  a_0 \Delta \Theta
_f^{\text{ NLO}}(s)  +
  a_0^2 \Delta \Theta_f^{\text{ NNLO}}(s)  \nonumber \\
\Delta \Theta_g
 & \equiv & \Delta \Theta_g^{\text{ LO}}(s)
 + a_0 \Delta \Theta
_g^{\text{ NLO}}(s) +
a_0^2 \Delta \Theta_g^{\text{ NNLO}}(s) \,,  \nonumber\\
\end{eqnarray}
one can finally rewrite the singlet distribution \eqref{eq:fsU-singlet} as
\begin{eqnarray}
\label{eq:la4}
U\Delta \mathcal{F}(s, U) &=&
\Delta \Phi_f \Delta \mathcal{F}
(s, U) + a_1 \Delta
 \Phi_f^{\text{NLO}} \Delta
 \mathcal{F}(s, U + b_1)
 + a_1^2 \Delta
 \Phi_f^{\text{NNLO}} \Delta
  \mathcal{F}(s, U + 2 b_1)
   + 2 a_0 a_1  \Delta
\Phi_f^{\text{NNLO}} \Delta
\mathcal{F}(s, U + b_1)  \nonumber  \\
&& + \Delta \Theta_f
\Delta \mathcal{G}(s, U) + a_1 \Delta \Theta
_f^{\text{NLO}} \Delta
 \mathcal{G}(s, U + b_1)
 + a_1^2 \Delta
\Theta_f^{\text{NNLO}} \Delta
\mathcal{G}(s, U + 2 b_1)
\nonumber \\
&& + 2 a_0 a_1 \Delta
 \Theta_f^{\text{NNLO}} \Delta
  \mathcal{G}(s, U + b_1)
 + \Delta f^0 \,,
\end{eqnarray}
and the gluon density \eqref{eq:gU-singlet} as
\begin{eqnarray}
\label{eq:la5}
U \Delta \mathcal{G}(s, U)
 &=& \Delta \Phi_g \Delta \mathcal{G}
(s, U) + a_1 \Delta
\Phi_g^{\text{NLO}} \Delta
\mathcal{G}(s, U + b_1)
 + a_1^2 \Delta
 \Phi_g^{\text{NNLO}} \Delta
  \mathcal{G}(s, U + 2 b_1)
 + 2a_0  a_1 \Delta
 \Phi_g^{\text{NNLO}} \Delta
  \mathcal{G}(s, U + b_1)  \nonumber
\\
&& + \Delta \Theta_g
\Delta \mathcal{F}(s, U) + a_1 \Delta \Theta
_g^{\text{NLO}} \Delta
 \mathcal{F}(s, U + b_1)
 + a_1^2 \Delta
\Theta_g^{\text{NNLO}}\Delta
 \mathcal{F}(s, U + 2 b_1)
\nonumber \\
&& + 2 a_0 a_1 \Delta
\Theta_g^{\text{NNLO}}\Delta
 \mathcal{F}(s, U + b_1)
 + \Delta g^0 \,.
\end{eqnarray}
\end{widetext}
The solution of Eqs.~\eqref{eq:la4} and \eqref{eq:la5} can be obtained via an iteration process.
We first consider the simple solutions of these equations which are labeled as $\Delta \mathcal{F}_{1}(s, U)$ and $\Delta \mathcal{G}_{1}(s, U)$.
These are concluded by setting $a_1 = 0$ and can be given by
\begin{eqnarray}
\label{eq:la6}
\lbrack U - \Delta \Phi_f (s)]
\Delta \mathcal{F}_1(s, U) - \Delta \Theta
_{f}(s) \Delta \mathcal{G}_1(s, U)
 &=& \Delta f_s^0 (s), \nonumber \\
-\Delta \Theta_g(s) \Delta
 \mathcal{F}_1 (s, U) + [U - \Delta \Phi
_g (s)] \Delta \mathcal{G}_1(s, U)
 &=& \Delta g^0 (s).  \nonumber\\
\end{eqnarray}
Their solutions lead to the following equations
\begin{eqnarray}
\label{eq:la7}
\Delta \mathcal{F}_1
&=& \frac{(U - \Delta \Phi_g)
	\Delta f_s^0 (s)}{
D(U, s)} + \frac {\Delta \Theta_f
 \Delta g^0(s)} {D(U, s)}, \nonumber\\
\Delta \mathcal{G}_1
&=& \frac{(U - \Delta \Phi_f)
	\Delta g^0(s)} {D(U, s)} +
\frac{\Delta \Theta_g  \Delta
	 f_s^0 (s)}{D(U, s)}.
\end{eqnarray}
The denominator $D(U, s)$ is the determinant of the coefficients of $\Delta \mathcal{F}(s, U)$ and $\Delta \mathcal{G}(s, U)$ in Eqs.~\eqref{eq:fsU-singlet} and \eqref{eq:gU-singlet}, i.e..
\begin{eqnarray}\label{eq:la8}
D(U, s) &=& \Delta \Phi_f(s)
\Delta \Phi_g (s) - \Delta \Theta_f (s)\Delta
\Theta_g(s)\nonumber \\
&&- [\Delta \Phi_f(s) +
 \Delta \Phi_g(s)] U + U^2
\end{eqnarray}

We now construct an iterative solution for the Eqs.~\eqref{eq:la4} and \eqref{eq:la5} to obtain the $\Delta \mathcal{F}$ and $\Delta \mathcal{G}$.
To achieve the first iteration solution, we need to change the arguments of $\Delta \mathcal{F}$ and $\Delta \mathcal{G}$ in Eq.~\eqref{eq:la7}.
They can be presented as $\Delta \mathcal{F}(s, U + b_{1})$, $\Delta \mathcal{G}(s, U + b_{1})$, $\Delta \mathcal{F}(s, U + 2 b_{1})$ and $\Delta \mathcal{G}(s, U + 2 b_{1})$, respectively.
The numerical values for the unknown parameters in Eq.~\eqref{eq:la3} at this order of iteration are extracted through a fitting procedure so we determined them as: $a_1 = 0.0591$, $b_1 = 7.0038$ and $a_0 = 0.00498$.
Now, to complete the calculations and to achieve the second step of the iteration process one can return from $U$ space to $\tau$ space using the inverse Laplace transform.
This yields the following expressions for the singlet and gluon distributions,
\begin{eqnarray}
\label{eq:la9}
\Delta f_s(s, \tau) &=&
k_{ff}(s, \tau)
\Delta f_s^0 (s) + k_{fg} (s, \tau
)\Delta g^0 (s), \nonumber\\
\Delta g(s, \tau)  &=&
k_{gg}(s, \tau )
\Delta g^0(s) + k_{gf}(s, \tau) \Delta
f_s^0(s).
\end{eqnarray}
The analytical expressions for the $k(s, \tau)$ coefficient functions, up to two step of iteration, are given in Appendix~\ref{AppendixB}. To obtain the spin-dependent PDFs as well as the structure functions in $x$ and ${\text Q}^2$ space, we used the numerical Laplace transform algorithms presented in Ref.~\cite{Block:2009en} for the numerical inversion
of Laplace transformations and convolutions.


%
%
\section{Leading-twist spin-dependent DIS structure function and Jacobi polynomials approach}\label{structure-function}

In the following, in detailed discussions we will describe the basic theoretical issues for a consistent determination of the spin-dependent PDFs from spin-dependent inclusive DIS data. We work in the framework of perturbative QCD at NNLO approximations using the $\overline{\rm MS}$ scheme for the renormalization and factorization.
In the leading twist (twist $\tau=2$) approximation, the spin-dependent proton structure function $g_1^p (s, {\text Q}^2)$ in Laplace $s$-space at NNLO  can be expressed as a linear combination of the spin-dependent PDFs and hard-scattering Wilson coefficient functions~\cite{Goto:1999by,Nath:2017ofm,Shahri:2016uzl,Khanpour:2017cha,Sato:2016tuz}, as
\begin{eqnarray}\label{eq:g1psspace}
&& g_1^{p,\,(\tau=2)} (s, {\text Q}^2) = \frac{1}{2}
\sum_q e^2_q  \Delta q_{v}
(s, {\text Q}^2)    \nonumber \\
&& \left(1 + \frac{\tau}{4 \pi}\Delta C^{(1)}_q +
\left(\frac{\tau}{4 \pi}\right)^2\Delta
 C^{(2)}_{ns}\right)      \nonumber   \\
&& + e^2_q (\Delta q_s+
\Delta \bar{q_s})(s, {\text Q}^2)  \nonumber \\
&&\left(1+\frac{\tau}{4 \pi}\Delta C^{(1)}_q
+ \left(\frac{\tau}{4 \pi}\right)^2
\Delta C^{(2)}_{s}\right)     \nonumber  \\
&& + e^2_q\Delta g(s, {\text Q}^2)
\left(\frac{\tau}{4 \pi}\Delta C^{(1)}_g
+\left(\frac{\tau}{4 \pi}\right)^2
\Delta C^{(2)}_g\right)\,.
\end{eqnarray}
In the above equation, $\Delta q_{v} (s, {\text Q}^2)$, $\Delta q_s (s, {\text Q}^2)$ and $\Delta g (s, {\text Q}^2)$ are the Laplace transforms of spin-dependent valance, sea and gluon densities, respectively.
The $\Delta C^{(1)}_q$ and $\Delta C^{(1)}_g$ are the NLO spin-dependent quark and gluon hard-scattering Wilson coefficients functions, calculable using the Laplace transform~\cite{AtashbarTehrani:2013qea}.
At NNLO approximation the Wilson coefficients are different for the quarks and antiquarks cases and we use $\Delta C^{(2)}_{ns}$ and $\Delta C^{(2)}_{s}$ calculated in Ref.~\cite{Zijlstra:1993sh}. The spin-dependent inclusive DIS is the only polarized process for which the hard-scattering coefficient functions are known up to NNLO.

In Ref.~\cite{Khanpour:2017cha}, we have shown that the power-suppressed ${{\cal O} (1/Q^2)}$ corrections to the structure functions as well as the target mass corrections (TMCs)  can make important contributions in some kinematic regions. In addition to the TMCs, spin-dependent structure functions in the operator product expansion (OPE) also receive contributions from higher twist terms (HTs) which are associated with matrix elements of multi-quark or quark-gluon operators.
In Ref.~\cite{Khanpour:2017cha}, we have also considered the contributions from the non-perturbative HTs. Since our main aim in this analysis is to study the applicability of Laplace transform to the QCD analysis of spin-dependent structure function, we restrict ourselves to the twist $\tau=2$ approximation in Eq.~\eqref{eq:g1psspace}.

The method employed in the {\tt SMKA18} QCD analysis is based on the Jacobi polynomials expansion of the spin-dependent structure functions. Practical aspects of this method including its major advantages are presented in our previous studies~\cite{Shahri:2016uzl,Khanpour:2017cha,MoosaviNejad:2016ebo,Khanpour:2016uxh} as well as other literature~\cite{Ayala:2015epa,Barker:1982rv,Barker:1983iy,Krivokhizhin:1987rz,Krivokhizhin:1990ct,Chyla:1986eb,Barker:1980wu,Kataev:1997nc,Alekhin:1998df,Kataev:1999bp,Kataev:2001kk,Kataev:2005ci,Leader:1997kw}.
Since the basic idea of Jacobi polynomials can be found in the mentioned references, we outline a brief review of this method. To illustrate this technique, we consider the case of $x g_1$ in Eq.~\eqref{eq:g1psspace}.
The spin-dependent structure function $x g_1(x, Q^2)$ is given by
\begin{eqnarray}\label{eq:xg1Jacobi}
x g_1(x, {\text Q}^2)  & =
 & x^{\beta}(1 - x)^{\alpha}
\, \sum_{n=0}^{N_{max}}
 \, \Theta_n^{\alpha, \beta}(x)
~ a_n({\text Q}^2)\,,
\end{eqnarray}
where, $n$ is the order of  expansion terms, $N_{\text {max}}$ is the number of polynomials which normally can be set to 7 or 9 and, $a_n({\text Q}^2)$ are the Jacobi moments.
The $\alpha$ and $\beta$ parameters are fixed as $\alpha=3$ and $\beta=0.5$, respectively. In our previous analyses, it is shown that with these selection one can achieve the fastest convergence of the above series~\cite{Shahri:2016uzl,Khanpour:2017cha,MoosaviNejad:2016ebo,Khanpour:2016uxh}.

One can conclude form Eq.~\eqref{eq:xg1Jacobi} that the use of Jacobi polynomials for the expansion of the structure functions has an advantage to allow one to factor out the essential part of the $x$-dependence of the structure functions into a weight function $x^{\beta}(1 - x)^{\alpha}$. Also, the ${\text Q}^2$-dependence is contained in the Jacobi moments.

The $\Theta_n^{\alpha, \beta}(x)$ in Eq.~\eqref{eq:xg1Jacobi} are the Jacobi polynomials with the following expansion,
\begin{equation}\label{eq:Jacobi-polynomials}
\Theta_n^{\alpha, \beta}(x) =
\sum_{j = 0}^{n} \, c_j^{(n)}
(\alpha, \beta) \, x^j \,,
\end{equation}
where $c_j^{(n)}(\alpha, \beta)$ are the coefficients which are expressed through $\gamma$-functions.
It is worth mentioning here that the Jacobi polynomials satisfy the following orthogonality relation with the weight function $x^{\beta} (1 - x)^{\alpha}$,
\begin{equation}\label{eq:orthogonality-relation}
\int_0^1 dx \, x^{\beta} (1 - x)^{\alpha}
\, \Theta_{k}^{\alpha, \beta}(x) \,
\Theta_{l}^{\alpha, \beta}(x) =
\delta_{k,l} \,.
\end{equation}
Using the above orthogonality relation, one can relate the spin-dependent proton structure functions with their Laplace $s$-space moments as follows
\begin{eqnarray}\label{eq:xg1Jacobii}
x g_1(x, {\text Q}^2)
& = & x^{\beta}(1 - x)^{\alpha}
\, \sum_{n=0}^{N_{max}}
\, \Theta_n^{\alpha, \beta}(x)
~ a_n({\text Q}^2)\, \nonumber \\
& = & x^{\beta}(1 - x)^{\alpha}
\, \sum_{n=0}^{N_{max}} \,
\Theta_n^{\alpha, \beta}(x) \nonumber \\
&\times &
 \sum_{j=0}^n \, c_j^{(n)}{(\alpha, \beta)}
\, {{\cal L}} [xg_1, s=j + 1] \,,
\end{eqnarray}
On the right-hand side of the above equation, the ${{\cal L}} [xg_1, s=j + 1]$ are the Laplace transformation of the structure functions.

%
%
\section { Overview of spin-dependent inclusive DIS datasets }\label{data}

In this section, we summarize the polarized DIS data sets used in {\tt SMKA18} QCD fits. We first review the analyzed data sets used in our work, then we will discuss on the present and future experimental efforts on hard-scattering polarized observables which can provide additional constrain on the gluon density and can also be used to separate the polarized quark and antiquark distributions.

The core of all spin-dependent PDFs fits include the spin-dependent DIS data obtained from neutral-current inclusive and semi-inclusive deep-inelastic scattering, DIS and SIDIS, with  charged lepton beams and nuclear targets at various fixed-target experiments. The set of  spin-dependent DIS data are provided in terms of proton, neutron and deuteron spin-dependent structure functions by different experiments at CERN, HERA and JLAB.
The polarized fixed-target experiment at HERMES (at HERA collider) have collected large amounts of data for the proton, neutron and deuteron as well as the heavier targets.
The {\tt SMKA18} global PDFs analysis uses all available data sets on the inclusive DIS of leptons over proton, neutron and deuteron that pass the required
cuts on the invariant final state mass $W^2 \geq 4 \, {\rm GeV}^2$ and $Q^2 \geq 1 \, {\rm GeV}^2$. This includes all of data sets from the EMC, SMC, COMPASS, SLAC, HERMES, and Jefferson Lab Hall A experiments which we also used in our previous global fits~\cite{Khanpour:2017cha,Shahri:2016uzl,Khanpour:2017fey}. Our analyzed data sets also include the most  recent high-precision data from {\tt COMPASS16} experiment at CERN~\cite{Adolph:2015saz,Adolph:2016myg}.

\begin{figure}[!htb]
\vspace*{0.40cm}
\includegraphics[clip,width=0.48\textwidth]{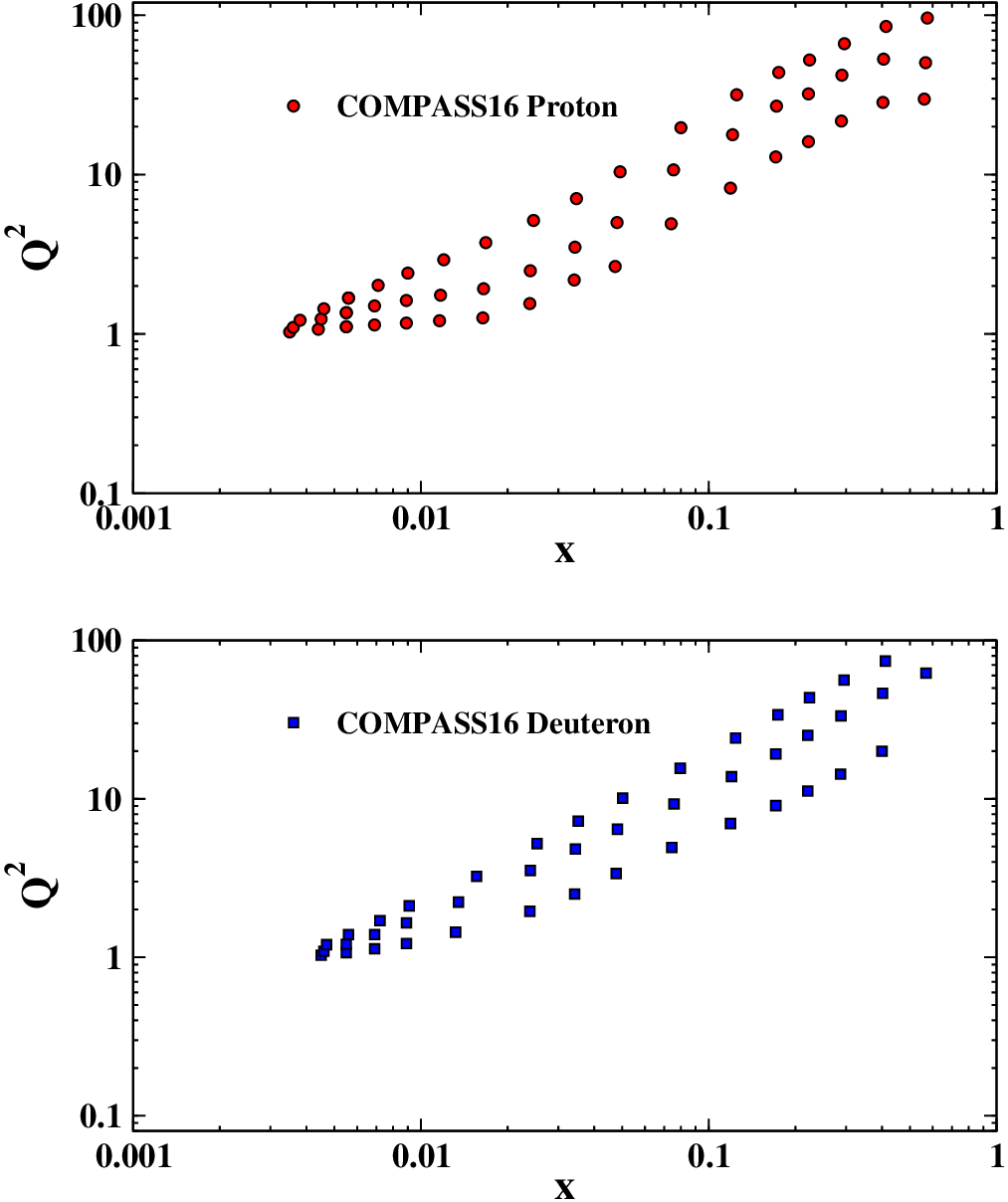}
\begin{center}
\caption{{\small (Color online) Representative kinematic coverage, in the (${\text Q}^2$, $x$) plane, of the most recent neutral current polarized inclusive DIS measurements on proton and deuteron targets reported by the {\tt COMPASS16} experiment at CERN~\cite{Adolph:2015saz,Adolph:2016myg}. These are used as input in {\tt SMKA18} global spin-dependent PDFs fit. \label{fig:COMPASS16}}}
\end{center}
\end{figure}

A list of published spin-dependent DIS experimental data points used in {\tt SMKA18} analysis are described in Table~\ref{tab:DISdata}. For each experiment we provided the $x$ and ${\text Q}^2$ kinematical ranges, the number of data points for each given target as well as the fitted normalization shifts ${\cal{N}}_i$ obtained from the QCD fit to the data.
As is seen from Table~\ref{tab:DISdata}, different combinations of spin-dependent inclusive DIS data sets for the proton, neutron and deuteron obtained by various experiments at CERN, HERA and JLAB are used in {\tt SMKA18} analysis. Our analysis does also contain the most recent data by {\tt COMPASS16}~\cite{Adolph:2015saz,Adolph:2016myg} experiment which are not used by other groups. In Refs.~\cite{Shahri:2016uzl} and \cite{Khanpour:2017cha}, we have shown that the inclusion of the recent {\tt COMPASS16} data leads to a reduction in the spin-dependent PDFs uncertainties for the valence and sea quarks as well as in the gluon PDFs uncertainty at the small value of $x$.

\begin{table*}[htb]
\caption{ Summary of published spin-dependent DIS experimental data points above ${\text Q}^2$ = 1.0 ${\text {GeV}}^2$ used in {\tt SMKA18} global analysis. For each experiment, it is listed the $x$ and ${\text Q}^2$ ranges, the number of data points for each given target, and the fitted normalization shifts ${\cal{N}}_i$ (see the text). } \label{tab:DISdata}
\begin{ruledtabular}
\begin{tabular}{l c c c c c c}
			\textbf{Experiment} & \textbf{Ref.} & \textbf{[$x_{\text {min}}, x_{\text {max}}$]}  & \textbf{${\text Q}^2$ range {(}${\text {GeV}}^2${)}}  & \textbf{\# data points} &   \textbf{${\cal N}_n$}   &     \tabularnewline
			\hline
			\text{\tt E143(p)}   & \cite{Abe:1998wq}   & [0.031--0.749]   & [1.27--9.52] & 28 & 1.000346&   \\
			\text{\tt HERMES(p)} & \cite{HERM98}  & [0.028--0.66]    & [1.01--7.36] & 39 & 1.001865&    \\
			\text{\tt SMC(p)}    & \cite{Adeva:1998vv}    & [0.005--0.480]   & [1.30--58.0] & 12 & 0.999911&    \\
			\text{\tt EMC(p)}    & \cite{EMCp}     & [0.015--0.466]   & [3.50--29.5] & 10 & 1.002207&    \\
			\text{\tt E155}      & \cite{E155p}    & [0.015--0.750]   & [1.22--34.72] & 24 & 1.024762&    \\
			\text{\tt HERMES06(p)} & \cite{Airapetian:2006vy} & [0.026--0.731]   & [1.12--14.29] & 51 &1.000182&     \\
			\text{\tt COMPASS10(p)} & \cite{Alekseev:2010hc} & [0.005--0.568]   & [1.10--62.10] & 15 &0.993010&    \\
			\text{\tt COMPASS16(p)} & \cite{Adolph:2015saz} & [0.0035--0.575]   & [1.03--96.1] & 54 & 1.000194&    \\
			\multicolumn{1}{c}{$\boldsymbol{g_1^p}(x,Q^2)$}       &  &  &  &  \textbf{233}  &&      \\
			\text{\tt E143(d)}  &\cite{Abe:1998wq}    & [0.031--0.749]   & [1.27--9.52]    & 28 &  0.999164&   \\
			\text{\tt SMC(d)}   &\cite{Adeva:1998vv}     & [0.005--0.479]   & [1.30--54.80]   & 12 &0.999988&   \\
			\text{\tt HERMES06(d)} & \cite{Airapetian:2006vy}& []0.026--0.731]   & [1.12--14.29]   & 51 & 0.998307&     \\
			\text{\tt E155(d)}  &\cite{E155d}     & [0.015--0.750]   & [1.22--34.79]   & 24 & 0.999915&    \\
			\text{\tt COMPASS05(d)}& \cite{COMP2005}& [0.0051--0.4740] & [1.18--47.5]   & 11 & 0.996924&    \\
			\text{\tt COMPASS06(d)}& \cite{COMP2006}& [0.0046--0.566] & [1.10--55.3]    & 15 & 0.999916&     \\
			\text{\tt COMPASS16(d)} & \cite{Adolph:2016myg} & [0.0045--0.569]   & [1.03--62.1] & 43 &1.000891&    \\
			\multicolumn{1}{c}{ $\boldsymbol{g_1^d} (x,Q^2)$}      &  &  & & \textbf{184} & &        \\
			\text{\tt HERMES(n)} &\cite{HERM98}   & [0.033--0.464]   & [1.22--5.25]    & 9 & 0.999958&    \\
			\text{\tt E154(n)}   &\cite{E154n}    & [0.017--0.564]   & [1.20--15.00]   & 17 & 0.999619&    \\
			\text{\tt HERMES06(n)} &\cite{Ackerstaff:1997ws}  &  [0.026--0.731]  & [1.12--14.29]   & 51 &1.000013&    \\
			\text{\tt JLAB03(n)}&\cite{JLABn2003} & ]0.14--0.22]     & [1.09--1.46]    & 4 & 0.999813&    \\
			\text{\tt JLAB04(n)}&\cite{JLABn2004} & [0.33--0.60]      & [2.71--4.8]     & 3 & 0.900000&    \\
			\text{\tt JLAB05(n)}&\cite{Kramer:2005qe} & [0.19--0.20]     &[1.13--1.34]     & 2 & 1.022321&    \\
			\text{\tt E142(n)}   &\cite{E142n}    & [0.035--0.466]   & [1.10--5.50]    & 8 & 0.998999&    \\
			\multicolumn{1}{c}{$\boldsymbol{g_1^n}(x,Q^2)$}     &     &  & & \textbf{94} & &     \\
			\multicolumn{1}{c}{\textbf{Total data points}}&\multicolumn{6}{c}{~~~~~~~~~~~~~~~~~~~~~~~~~~~~~~~~~~~~~~~~~~~~~~~~~\textbf{511}}  			\\
\end{tabular}
\end{ruledtabular}
\end{table*}

The kinematic coverage of recent {\tt COMPASS16} data  for the proton and deuteron targets~\cite{Adolph:2015saz,Adolph:2016myg}  is displayed in Fig.~\ref{fig:COMPASS16}. These are used in our analysis to set an additional constrain on the spin-dependent PDFs. From this figure, one  can conclude  that the quantity of high precision data points from {\tt COMPASS16} experiments  and their kinematic coverage are presently much more limited in the polarized case. Therefore, the polarized PDFs can currently be determined with much less precision than the unpolarized PDFs and only cover the $x$-range for $x \geq 0.0035$.
One can expect that the future LHC data will certainly represent important opportunities to improve our knowledge on the spin-dependent PDFs. However, there are many challenges for the spin-dependent PDFs fitters to include such increasingly precise datasets effectively within a spin-dependent PDFs fit. A summary  of these data sets can be found in Refs.~\cite{Khanpour:2017cha,Shahri:2016uzl,Khanpour:2017fey,Lin:2017snn,Bourrely:2015kla}.

In addition to the DIS and SIDIS fixed-target data sets mentioned above, a significant amount of spin-dependent data from longitudinally polarized proton-proton  collisions at {\tt RHIC} has become available recently, which is however, in a limited range of momentum fractions, $0.05 < x < 0.4$~\cite{Aschenauer:2015eha}.
The longitudinal single-spin and double-spin asymmetries for the weak boson productions  are sensitive to the flavor decomposition. These can also be used to separate the polarized quark and antiquark distributions~\cite{Bourrely:1993dd}.

The STAR data sets on the $W$-boson production at polarized proton-proton collisions at RHIC provide evidence of a positive $\Delta \bar u$ distribution and a negative $\Delta \bar d$ distribution~\cite{Adamczyk:2014xyw}. The double-spin asymmetries for the production of jet, di-jet and $\pi^0$ are also  sensitive to the gluon polarization~\cite{Bourrely:1990pz}, directly. The kinematic coverage of the spin-dependent data, the quantity of the data points, and the variety of available hard-scattering processes are presently much more limited for the polarized case in comparison with the unpolarized one~\cite{Lin:2017snn}. Hence, the spin-dependence PDFs can currently be determined with much less precision than the unpolarized PDFs, specially at the small range of $x$.
The kinematic coverage is expected to be significantly improved in the future, with the DIS and SIDIS data from 12 GeV upgrade of Jefferson Lab~\cite{Dudek:2012vr} and future spin-dependent high-energy and high-precision Electron-Ion Collider (EIC)~\cite{Accardi:2012qut}.
We should mention here that the QCD analyses of spin-dependent nuclear-target data requires an accurate account of nuclear corrections. For recent reviews see, for example, Refs.~\cite{Khanpour:2017fey,Flay:2016wie,Yan:2016ods,Guzey:2000wh,Frankfurt:1996nf,Guzey:1999rq,Ethier:2013hna}.
Briefly, we have demonstrated that in the spin-dependent case, even with current uncertainties, new and precise data from JLAB, RHIC, EIC and CERN can impose sizable constraints on several important spin-dependent quark combinations. This suggests, the  global spin-dependent PDFs analyses should include future data sets in their fits to  constrain the gluon density much more. These new data can also constrain   some of the less well-known quark combinations, such as the total strangeness.

%
%
\section{ Spin-dependent PDFs parametrizations and errors }\label{pdf}

We consider a proton comprised of massless partons with spin-dependent distributions $q_\pm(x, Q^2)$ which carry the momentum
fraction of $x$ with a characteristic scale ${\text Q}$. In the present analysis, for the generic parametrization of the spin-dependent PDFs, we take into account the following standard functional form at the initial scale $Q_0^2=1 \, \text {GeV}^2$,
\begin{equation}
\label{eq:partonQ0}
x \Delta q(x, Q_{0}^{2}) =
{\cal N}_q \eta_q x^{a_q}(1 - x)^{b_q}
(1 + c_{q} x) \,,
\end{equation}
which includes four shape parameters $\eta_q$, $a_q$, $b_q$, $c_q$, and the normalization coefficients ${\cal N}_q$.
The generic labels $q = {u_v, d_v, \overline{q}, g}$ refer to the partonic flavors of up-valence, down-valence, sea, and gluon, respectively.
The normalization constants ${\cal N}_q$ are determined as
\begin{equation}\label{eq:normal}
\frac{1}{ {\cal N}_{q}} =
\left(1 + c_{q} \frac{a_{q}}{1 + a_{q} + b_{q}} \right)
 B(a_{q}, b_{q} + 1),
\end{equation}
where the function ${\rm B}(a, b)$ is the Euler $\beta$ function.
These are chosen such that $\eta_{q}$ are the first moments of $x \Delta q(x, Q_0^2)$. Since the present analysis just considers the spin-dependent inclusive DIS data, then we attempt to assume an ${\rm SU}(3)$ flavor symmetry such that $\Delta \overline{q} (x, Q^{2}) \equiv \Delta \overline{u} (x, Q^{2}) = \Delta \overline{d}  (x, Q^{2}) = \Delta \overline{s} (x, Q^{2}) =\Delta s (x, Q^{2})$. Therefore, we try to fit only the spin-dependent PDFs of $x \Delta u_v$, $x \Delta d_v$, $x \Delta \overline{q}$ and $x \Delta g$.

One can also consider additional constraints on the moments of the spin-dependent PDFs which could provide by the weak neutron and hyperon decay constants.
Hence, the first moments of the spin-dependent valence distribution can be described in terms of the axial charges for octet baryons, ${\text F}$ and ${\text D}$, in which measured in hyperon and neutron $\beta$ decay~\cite{Bass:2009ed}. These constraints lead to the values of $\eta_{u_v} = 0.928 \pm 0.014$ and $\eta_{d_v} = -0.342 \pm 0.018$~\cite{Agashe:2014kda}. We fix two valence first moments on their central values. The parameters $\eta_{\overline{q}}$ and $\eta_{g}$ are determined from the QCD fit to inclusive data.

For the spin-dependent quark and gluon distributions in Laplace $s$-space, we follow our previous PDFs analyses~\cite{Khanpour:2016uxh,MoosaviNejad:2016ebo}. As we discussed in Sec.~\ref{structure-function}, in the Jacobi polynomial approach, the DGLAP evolution equations can also be solved in the Laplace $s$-space. The Laplace transformation of the spin-dependent PDFs $\Delta q$ are defined as in Eq.~\eqref{eq:partonQ0}, i.e.
\begin{eqnarray}
{\cal L}[\Delta q(x& = &e^{-v}, Q_0^2),s]
\equiv \Delta q(s,Q_0^2) \nonumber \\
&=& \int_{0}^{\infty}e^{- sv}
\Delta q(x = e^{-v}, Q_0^2) dv  \\
&=& {\cal N}_q\eta_q
\left(1 + c_q\frac{s + a_q}
{1 + a_q + b_q + s} \right)
B(s + a_q, b_q + 1)\,. \nonumber
\end{eqnarray}
The fit parameters are determined by minimizing the $\chi^2_{\rm global} (\{\eta_k\}) $ function, which are defined as
\begin{eqnarray}\label{chi2}
\chi^2_{\rm global} (\{\eta_k\}) =
\sum_{i=1}^{n^{exp}} \bigg[ \sum_{j=1}^{N^{data}}&&
\bigg( \frac{D_j^i {{\cal N}_j^i}
- T_j^i (\{\eta_k\}) }
{\sigma_j^i  {{\cal N}_j^i}}  \bigg)^2 + \nonumber \\
 &&( \frac{1 - {\cal N}_i }  { \Delta {\cal N}_i })^2 \bigg] \,,
\end{eqnarray}
where $\{\eta_k\}$ is a set of fitted parameters, $D_j^i$ is the measured value of the observable for the data point $j$ from the experimental data set $i$.
In Eq.~\eqref{chi2}, $T_j^i(\{\eta_k\})$ is the corresponding theoretical value and $\sigma_j^i$ represents the uncorrelated statistical and systematic uncertainties added in quadrature.
Another important issue that is addressed in {\tt SMKA18} global analysis is the estimation of uncertainties in the extraction of various spin-dependent PDFs, associated with experimental uncertainties. {\tt SMKA18} pursues here an approach based on the use of the Hessian method~\cite{Pumplin:2001ct}.
In the next section, we will show that in the standard single-fit PDF analyses, one often finds that some of the shape parameters in Eq.~\eqref{eq:partonQ0} are not well determined by spin-dependent inclusive data and need to be fixed by hand.

\begin{figure*}[!htb]
\vspace*{0.40cm}
\includegraphics[clip,width=0.60\textwidth]{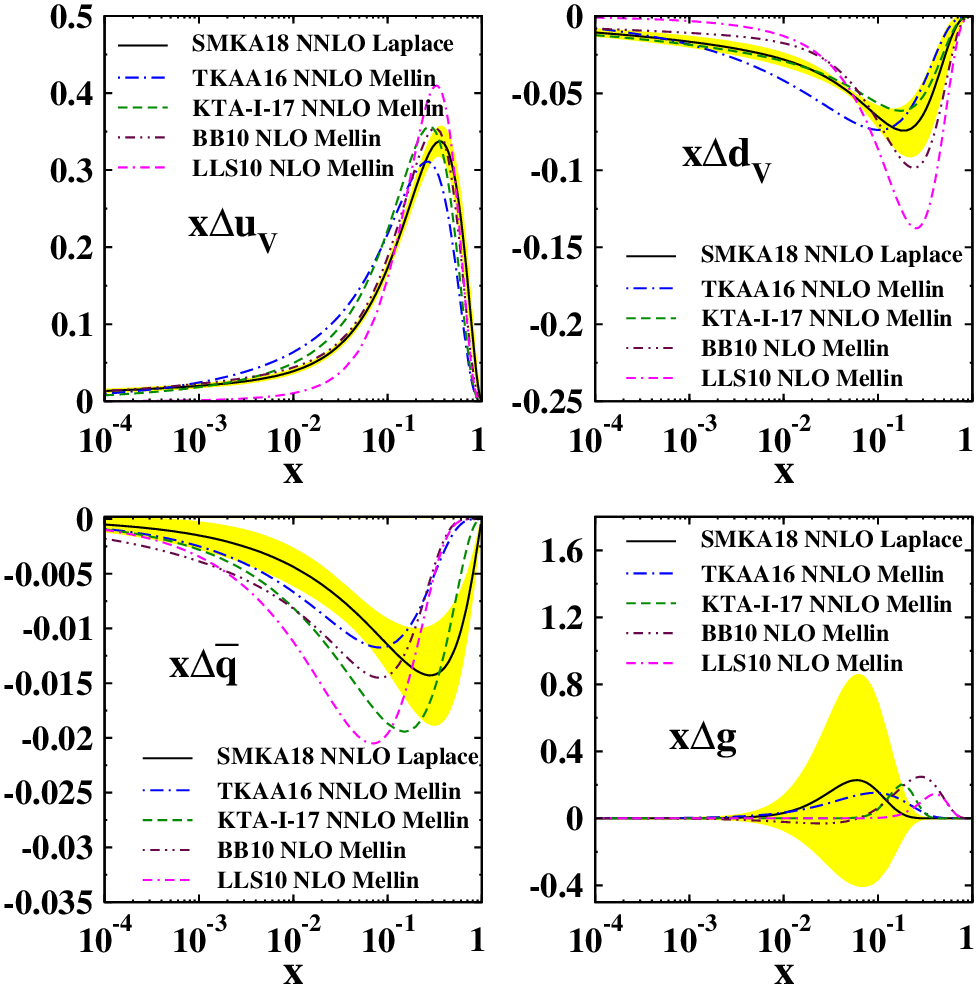}
\begin{center}
\caption{{\small (Color online) {\tt SMKA18}  spin-dependent PDFs as a function of $x$ at the input scale $Q_0^2 = 1 \, {\rm GeV}^2$. The Mellin space results from {\tt KTA-I-17}~\cite{Khanpour:2017cha} (dashed) and {\tt TKAA16}~\cite{Shahri:2016uzl} (dashed-dotted)  at NNLO accuracy in pQCD are also shown.  The NLO results from {\tt BB10}~\cite{Blumlein:2010rn} and {\tt LLS10}~\cite{Leader:2010rb} are  shown as well. \label{fig:xg1pnnloQ0}}}
\end{center}
\end{figure*}

\begin{figure}[!htb]
\vspace*{0.40cm}
\includegraphics[clip,width=0.45\textwidth]{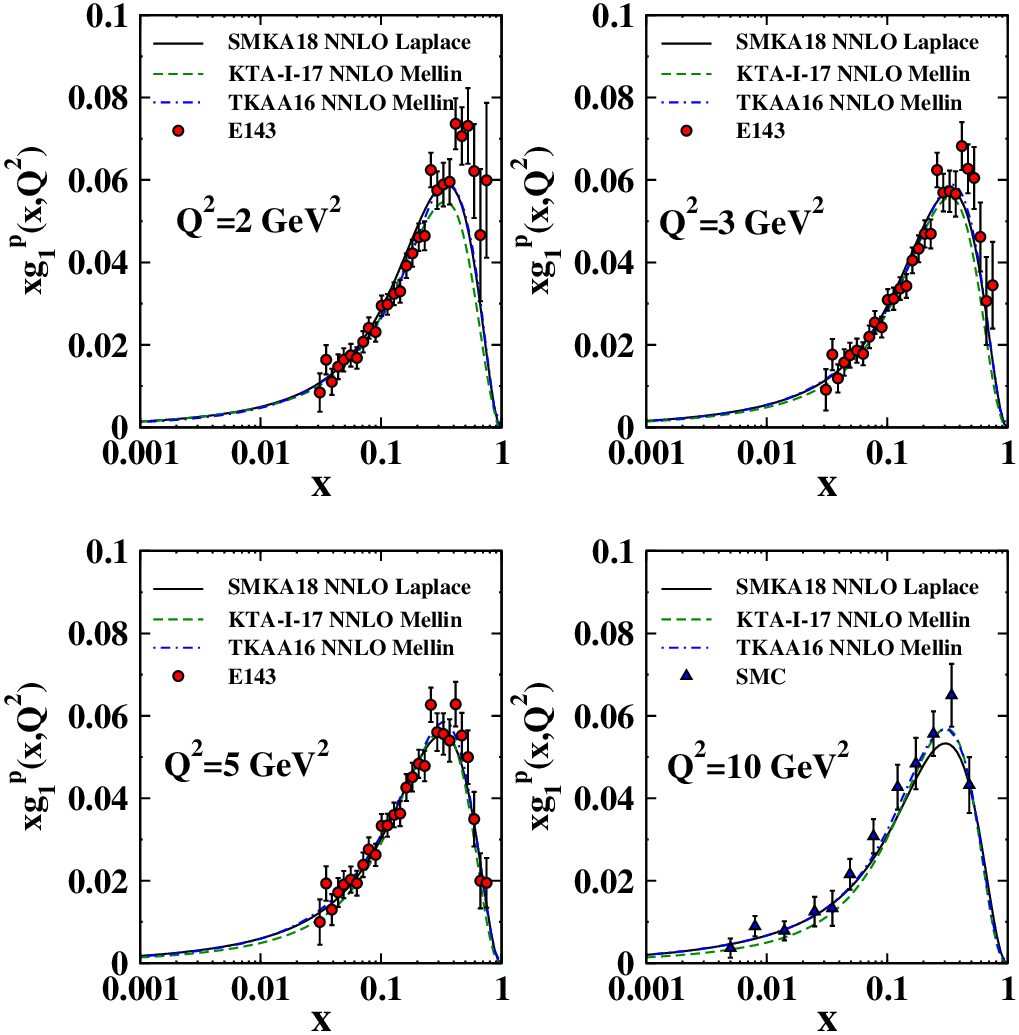}
\begin{center}
\caption{{\small (Color online)  The results of {\tt SMKA18} NNLO pQCD fit for the spin-dependent proton structure functions  are shown as a function of $x$. The results from Mellin space analyses of {\tt KTA-I-17}~\cite{Khanpour:2017cha} (dashed) and {\tt TKAA16}~\cite{Shahri:2016uzl} (dashed-dotted) are also  shown. \label{fig:xg1pnnlo}}}
\end{center}
\end{figure}

\begin{figure}[!htb]
\vspace*{0.40cm}
\includegraphics[clip,width=0.40\textwidth]{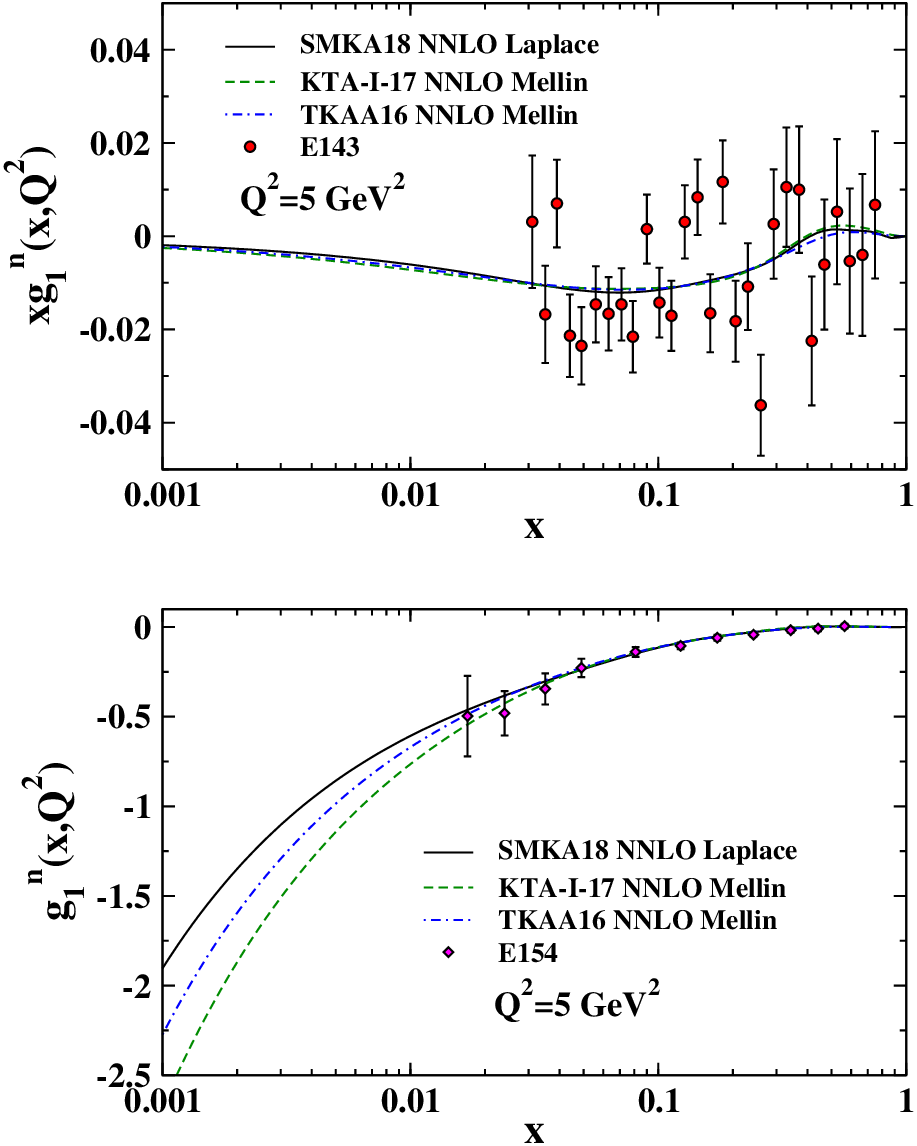}
\begin{center}
\caption{{\small (Color online) As in Fig.~\ref{fig:xg1pnnlo}, but for the neutron structure functions.
 \label{fig:xg1nnnlo}}}
\end{center}
\end{figure}

\begin{figure}[!htb]
\vspace*{0.40cm}
\includegraphics[clip,width=0.40\textwidth]{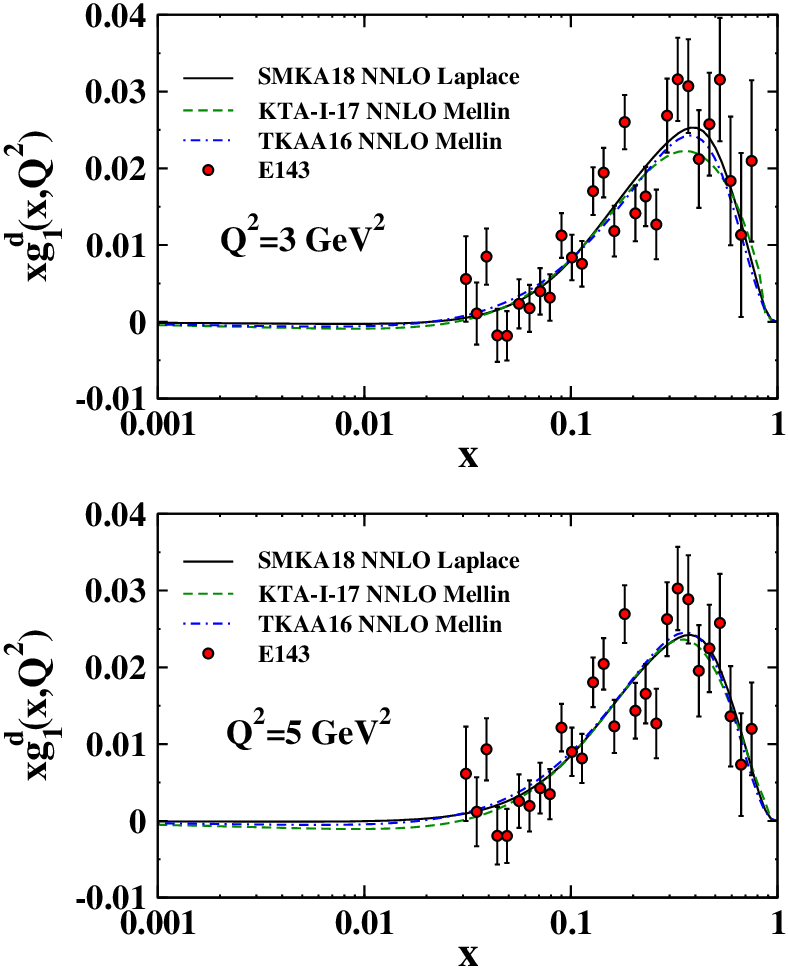}
\begin{center}
\caption{{\small (Color online) As in Fig.~\ref{fig:xg1pnnlo}, but for the deuteron structure functions.
 \label{fig:xg1dnnlo}}}
\end{center}
\end{figure}

%
%
\section{ Results and Discussions }\label{result}

 In this section, we will present our detailed discussions on the obtained results and compare our extracted results with the others in  literature.
Having our methodology for fitting the spin-dependent PDFs, which includes the spin-dependent structure functions in Laplace space, the analyzed observables and the {\tt SMKA18} spin-dependent input,  we are now in a position to present our main results of {\tt SMKA18} analysis in this Section.

Before starting our detailed discussions on {\tt SMKA18} spin-dependent PDF determinations, we would like to illustrate the present status of spin-dependent PDFs sets at NLO and NNLO QCD approximations and discuss some differences, briefly, which are clearly visible.
The available PDF sets at NLO and NNLO approximations are listed in Table.~\ref{tab:PDFsets} which include {\tt AAC09}~\cite{Hirai:2008aj}, {\tt BB10}~\cite{Blumlein:2010rn}, {\tt  LSS10}~\cite{Leader:2010rb}, {\tt NNPDF14}~\cite{Nocera:2014gqa}, {\tt JAM13~\cite{Jimenez-Delgado:2013boa}}, {\tt JAM14}~\cite{Jimenez-Delgado:2014xza}, {\tt JAM15}~\cite{Sato:2016tuz} at NLO, and {\tt TKAA16}~\cite{Shahri:2016uzl}, {\tt KTA-I-17}~\cite{Khanpour:2017cha} and {\tt KTA-II-17}~\cite{Khanpour:2017fey} at NNLO approximations.
In the first column of Table.~\ref{tab:PDFsets}, we indicate the name of the group and in the subsequent columns we present the theory accuracy, the data sets and the corresponding references.
These spin-dependent PDFs determinations are different in the input parametrizations, the datasets included in the analyses, the details of the QCD analysis such as the theory accuracy and the treatment of HT and TMCs, and finally in the procedure used to determine the spin-dependent PDFs from the data.
In the theory setup point of view, the DSSV has developed a method based on ``Mellin moments'' of the PDFs and JAM collaboration has implemented a new approach called iterative ``Monte Carlo'' procedure~\cite{Sato:2016tuz,Ethier:2017zbq}.
The {\tt DSSV14}~\cite{deFlorian:2014yva} and {\tt NNPDFpol1.1}~\cite{Nocera:2014gqa} spin-dependent PDFs update their previous analyses, {\tt DSSV08}~\cite{deFlorian:2009vb,deFlorian:2008mr} and {\tt NNPDFpol1.0}~\cite{Ball:2013lla} have included almost all available experimental information. The impact of RHIC proton-proton data have been studied by the inclusion of data from double-spin asymmetries for inclusive jet production and $\pi^0$ production~\cite{Adamczyk:2014ozi,Adare:2014hsq} and the data on double-spin asymmetries for high-$p_T$ inclusive jet production~\cite{Adamczyk:2014ozi,Adamczyk:2012qj,Adare:2010cc} and weak boson production~\cite{Adamczyk:2014xyw}.

\begin{table*}[htb]
\begin{tabular}{cccccc}
\hline
Polarized PDF sets  &  Theory accuracy  &  Data sets                                        &  &   Ref.                       \\   \hline \hline
{\tt NNPDF14}       &  NLO              &  Asymmetry \& double Asymmetry \& Polarized DIS   &  & \cite{Nocera:2014gqa}        \\
{\tt AAC09}         &  NLO              &  Asymmetries                                      &  &  \cite{Hirai:2008aj}         \\
{\tt BB10}          &  NLO              &  Asymmetries \& Polarized DIS                     &  &  \cite{Blumlein:2010rn}      \\
{\tt LSS10}         &  NLO              &  Polarized DIS \& SIDIS                           &  &  \cite{Leader:2010rb}        \\
{\tt DSSV14}        &  NLO              &  Polarized DIS \& SIDIS \& $pp$                   &  &  \cite{deFlorian:2014yva}    \\
{\tt JAM13}         &  NLO              &  Asymmetries                                      &  &  \cite{Jimenez-Delgado:2013boa}    \\
{\tt JAM14}         &  NLO              &  Asymmetries                                      &  &  \cite{Jimenez-Delgado:2014xza}    \\
{\tt JAM15}         &  NLO              &  Asymmetries                                      &  & \cite{Sato:2016tuz}                \\
{\tt TKAA16}        &  NNLO             &  Polarized DIS                                    &  &  \cite{Shahri:2016uzl}        \\
{\tt KTA-I-17}      &  NNLO             &  Polarized DIS                                    &  &  \cite{Khanpour:2017cha}      \\
{\tt KTA-II-17}     &  NNLO             &  Polarized DIS                                    &  &  \cite{Khanpour:2017fey}      \\
\hline \hline
\end{tabular}
\caption{ Current status of the most recent analyses of spin-dependent PDFs. \label{tab:PDFsets}  }
\end{table*}

The data from {\tt COMPASS16} experiments at LHC~\cite{Adolph:2015saz,Adolph:2016myg} certainly represent important opportunities to improve the knowledge on spin-dependent PDFs. There are, however, many challenges for the  spin-dependent PDFs fitters to include such increasingly precise data effectively within a spin-dependent PDFs fit. These data can be used to further constrain the spin-dependent PDFs and to measure the strong coupling constant $\alpha_{s}$. These experiments cover a wide range of physical observables. Additional experimental information is expected from ongoing and future experimental efforts.   {\tt COMPASS16} data sets on polarized proton and deuteron targets have been used in our previous NNLO QCD analyses. As we mentioned, in our present analysis we also use these precise data sets. Hence, our main aim of this paper is to introduce a new method to extract the spin-dependent PDFs; the Laplace transform and Jacobi polynomials approach.

After this brief review on recent efforts on spin-dependent PDFs analyses, we now present the results obtained for the {\tt SMKA18} spin-dependent PDFs at NNLO in the basis of the Laplace transform technique and Jacobi polynomial approach. The {\tt SMKA18} fit parameters are presented in Table.~\ref{tab:dataparton}. The parameters $c_{\overline{q}}$ and $c_{g}$ are set to zero because the current spin-dependent inclusive DIS data can not constrain all the fit parameters. As we have mentioned earlier, despite of the outstanding achievements, at the present the spin-dependent PDFs cannot be well determined in a global QCD analysis with a high accuracy as one has for the unpolarized ones. The experimental data on spin-dependent DIS are relatively span narrow range of $x$ and ${\text{Q}}^2$. As a consequence, the quarks, antiquarks, and gluon densities are still affected by large uncertainties.

Now, we turn to discuss in more detail , how the results on {\tt SMKA18} spin-dependent PDFs depend on the method used for their determination. To illustrate this dependence, we will compare {\tt SMKA18} NNLO set of spin-dependent PDFs determined by Laplace method with those obtained by {\tt KTA-I-17}~\cite{Khanpour:2017cha} and {\tt TKAA16}~\cite{Shahri:2016uzl} using the Mellin space analyses.  The differences between the mentioned NNLO spin-dependent PDFs lie in the formalism used in these analyses, Mellin and Laplace space, as well as the treatment of higher twist and TMCs.

The final distributions for the {\tt SMKA18} fit are displayed in Fig.~\ref{fig:xg1pnnloQ0} as a function of $x$ at fixed  $Q_0^2 = 1 \, \text {GeV}^2$.
For comparison, we have also shown the Mellin space results from recent NNLO analyses of {\tt KTA-I-17}~\cite{Khanpour:2017cha} (dashed) and {\tt TKAA16}~\cite{Shahri:2016uzl} (dashed-dotted).  In order to see the effect of higher order corrections and its comparison with the NLO analyses, the {\tt BB10}~\cite{Blumlein:2010rn} and {\tt LLS10}~\cite{Leader:2010rb} results for the spin-dependent PDFs at NLO pQCD accuracy are also shown  in Fig.~\ref{fig:xg1pnnloQ0}.
 As one can see from this figure,  the $x \Delta u_{v}$ and $x \Delta d_{v}$ PDFs are the best determined distributions from the spin-dependent inclusive DIS data, with relatively small uncertainty bands.
We should stress here that the uncertainties are computed using the ``Hessian method''. The contributions from the extrapolated regions, $x < 0.001$ and $x > 0.8$, the spin-dependent PDFs are not directly constrained by the inclusive DIS datasets.
For the much better determined $x \Delta u_{v}$ and $x \Delta d_{v}$ distributions, one can conclude that the shapes and magnitudes from the {\tt SMKA18} QCD fit are generally similar to those found in our previous Mellin based analysis~\cite{Khanpour:2017cha,Shahri:2016uzl}.  In comparison with the available NLO analyses, one can see that our NNLO results are compatible with the {\tt BB10}~\cite{Blumlein:2010rn} while the {\tt LLS10}~\cite{Leader:2010rb} shows higher spin-dependent valence distributions.

The strange quark distribution $x \Delta \bar q$ turns out to be negative. In contrast to the negative sea quark distribution obtained from {\tt SMKA18} analysis of inclusive DIS, {\tt  DSSV09}~\cite{deFlorian:2009vb} and {\tt LSS10}~\cite{Leader:2010rb} fits have shown that the inclusion of the semi-inclusive kaon production data in the QCD fit induces a positive  value at the Bjorken value of $x > 0.05$. It is constrained by a combination of ${\text Q}^2$ evolution, the weak baryon decay constants, and the assumption of an $\text {SU}(3)$ symmetric sea, $\Delta \overline{q} (x, Q^2) \equiv \Delta \overline{u} (x, Q^2) = \Delta \overline{d}  (x, Q^2) = \Delta \overline{s} (x, Q^2) =\Delta s (x, Q^2)$.
A much wider error band has been obtained for the sea quark distribution, as is shown in Fig.~\ref{fig:xg1pnnloQ0}.  As a function of $x$, the shape of the {\tt SMKA18}  $x \Delta \bar q$ PDFs is slightly smaller than the one from {\tt  KTA-I-17}~\cite{Khanpour:2017cha} and larger than the one from {\tt TKAA16}~\cite{Shahri:2016uzl}, which are extracted from the Mellin based analysis.  Both the {\tt BB10}~\cite{Blumlein:2010rn} and the {\tt LLS10}~\cite{Leader:2010rb} strange quark densities are larger then all of our NNLO strange quark densities.

From Fig.~\ref{fig:xg1pnnloQ0}, it is also seen that for the gluon distributions there is a  much wider error band than all other spin-dependent quark PDFs. This wide uncertainty is expected to be reduced when the pion and jet production data from spin-dependent proton-proton collisions are included in the analysis~\cite{Sato:2016tuz}.
The difficulty in constraining the spin-dependent $x\Delta g(x, Q^2)$ distribution is clearly revealed through the spread of gluon distribution from various global PDFs parametrizations. In Fig.~\ref{fig:xg1pnnloQ0}, the $x\Delta g(x, Q^2)$ PDFs from our previous NNLO global analyses are compared with the {\tt SMKA18} results. Note that, the {\tt KTA-I-17} and {\tt TKAA16} QCD fits have used spin-dependent inclusive DIS data, similar to the present analysis.
As is seen, in our all fits the $x\Delta g$ PDFs are positive in all range of $x$.  The {\tt LLS10} gluon density approaches zero faster than the other curves and the {\tt BB10} shows a sign change at the medium value of $x$.
In most of the QCD fits the gluon distributions is positive at large values of $x$, with a sign change at smaller $x$ values. However, the latest analyses by {\tt  DSSV14}~\cite{deFlorian:2014yva} and {\tt JAM15}~\cite{Sato:2016tuz} of the recent high-statistics jet data from RHIC does also give a positive $x \Delta g$ distribution with no indication of a sign change in the measured $x$ region.

In the rest of this section, we compare the results of our analysis with the analyzed inclusive spin-dependent DIS data. In Fig.~\ref{fig:xg1pnnlo}, we presented a detailed comparison of {\tt SMKA18} NNLO pQCD fit results  for the spin-dependent structure functions with the analyzed inclusive DIS data for some selected values of ${\text Q}^2$ = 2, 3, 5 and 10 $\text {GeV}^2$. The results from Mellin space analyses of {\tt KTA-I-17}~\cite{Khanpour:2017cha} (dashed) and {\tt TKAA16}~\cite{Shahri:2016uzl} (dashed-dotted)  have been also shown for comparison. As one can see, for all these spin-dependent PDFs sets, the perturbative expansion is stable and the obtained theory computations agree well with the spin-dependent inclusive DIS data for all range of $x$. The same comparison for the spin-dependent proton and neutron structure functions are displayed in Figs.~\ref{fig:xg1nnnlo} and \ref{fig:xg1dnnlo}. Interestingly, the same pattern is also seen for the neutron and deuteron, and the {\tt SMKA18} theory predictions are in good agreement with all DIS data both at low and high values of $x$.

As a short summary, we have performed a new global QCD analysis of spin-dependent PDFs including all available inclusive DIS data from experiments at CERN, SLAC, HERA and JLAB.
The analysis is the first QCD fit at NNLO performed using a developed strategy based on Laplace transform technique. One can conclude that our findings for the theory predictions of spin-dependent structure functions are in good agreements with the analyzed inclusive DIS data. These results indicate the validity of the Jacobi polynomials and Laplace technique for the case of spin-dependent QCD analyses.

\begin{table}[H]
	\begin{tabular}{cccccc}
		\hline\hline
		\multicolumn{6}{c}{\text{ NNLO} } \\ \hline
		$\Delta u_{v}$ & $\eta _{u_{v}}$      & $0.928$ (fixed) & $\Delta \overline{q}$ & $%
		\eta _{\overline{q}}$ & $-0.05305 \pm 0.00894$    \\
		& $a_{u_{v}}$ & $0.1606 \pm 0.0295$   &  & $a_{\overline{q}}$ & $0.4693 \pm 0.191$  \\
		& $b_{u_{v}}$ & $1.973 \pm 0.126$     &  & $b_{\overline{q}}$ & $1.1915 \pm 0.199$   \\
		& $c_{u_{v}}$ & $42.978 \pm 7.831$    &  & $c_{\overline{q}}$ & $0^*$   \\
		$\Delta u_{v}$ & $\eta _{d_{v}}$      &   $-0.342$ (fixed) & $\Delta g$ & $\eta _{g}$ & $0.4065\pm 0.955$   \\
		& $a_{d_{v}}$ & $0.1918 \pm 0.007024$ &  & $a_{g}$ & $2.0159 \pm 5.47$  \\
		& $b_{d_{v}}$ & $4.0749 \pm 0.614$    &  & $b_{g}$ & $31.929 \pm 28.53$    \\
		& $c_{d_{v}}$ & $15.416 \pm 5.6$      &  & $c_{g}$   & $0^*$  \\
		\multicolumn{6}{c}{$\Lambda = 0.2043 \pm 0.0145 \, \text{GeV}$}   \\
		\multicolumn{6}{c}{$\chi^2/{\rm d.o.f} = 459.898/498 = 0.923$}  \\ \hline \hline
	\end{tabular}
	\caption{ Final parameter values and their statistical errors in the $\overline{\rm MS}$ scheme at the input scale $Q_0^2  = 1  \, \text {GeV}^2$, in the NNLO approximations. \label{tab:dataparton} }
\end{table}

%
%
\section{Summary and Conclusions}\label{Summary}

The detailed study of the spin-dependent PDFs of the nucleon and the spin structure of hadron are active interdisciplinary research field lying at the crossroads of
high-energy hadronic and nuclear physics, with important applications in LHC phenomenology. In this paper, our main aim was to maximally utilize the most available spin-dependent inclusive DIS data over the greatest range of kinematics considering the theoretical perturbative QCD in Laplace $s$-space. The main outcome of this paper is the first quantitative study of the method of Laplace transform technique in the global QCD fits, for the spin-dependent case.
While {\tt SMKA18} study should and will be improved on a number of aspects, in particular related to the inclusion of new spin-dependent inclusive DIS data or treatment of TMCs and HT effects, we believe that it opens a new method on the determination of spin-dependent structure of the nucleon.
Finally, note that the use of our Laplace technique and Jacobi polynomials approach for incorporating NNLO theory accuracy in the analysis has a number of important implications for future practice and can be extended in a number of directions. Overall, very good descriptions of the global spin-dependent inclusive DIS data set has been obtained in {\tt SMKA18} QCD fit, over the entire range of $x$ and ${\text Q}^2$ covered by the preferred cuts.
In the near future, the data from 12 ${\rm GeV}$ {\tt JLAB} experiments will provide stronger constraints on the behavior of spin-dependent PDFs at large $x$ through precise measurements of polarization asymmetries over a greater range of ${\text Q}^2$ and ${\text W}^2$~\cite{Sato:2016tuz,Jimenez-Delgado:2014xza,Jimenez-Delgado:2013boa}.
In addition, the inclusion of semi-inclusive DIS asymmetries as well as jet and pion production asymmetries in spin-dependent proton-proton ($pp$) collisions, will place stronger constraints on the sea quark polarization~\cite{Sato:2016tuz}.
Finally, it is pointed out that the very accurate spin-dependent DIS data in a wide region require a more careful matching of QCD to the data in order to determine the spin-dependent PDFs correctly.

%
%
\section*{Acknowledgments}

The Authors acknowledge Oleg V. Selyugin for the useful discussions and detailed comments on the manuscript.
We are thankful to the School of Particles and Accelerators, Institute for Research in Fundamental
Sciences (IPM) for financial support of this project. S. M. Moosavi Nejad thanks the Yazd University for financial support provided for this project.
Hamzeh Khanpour also gratefully acknowledge the University of Science and Technology of Mazandaran for financial support provided for this research.


\appendix

%

%
\begin{widetext}
\section{NNLO splitting functions at NNLO approximation }\label{AppendixA}

Here, we present the Laplace transforms of the NNLO splitting functions~\cite{Moch:2014sna} for nonsinglet and singlet sectors, denoted
by $\Delta \Phi_{\text {NS}+}^{\text {NNLO}}$, $\Delta \Phi_{\rm ps}^{\text {NNLO}}$, $\Delta \Theta_f^{\text {NNLO}}$, $\Delta \Theta_g^{\text {NNLO}}$ and $\Delta \Phi_g^{\text {NNLO}}$, where $\gamma_E = 0.577216$ is the Euler constant, $\psi(n) = d\ln\Gamma(n)/dn$ is the digamma function and $\zeta (3) = 1.20206$.

\begin{eqnarray*}
&&\Delta \Phi _{ps}^{\rm NNLO}=\hspace{14 cm}\text{(A.1)}\\
&&f^2 \left[-\frac{7.0446}{(1+s)^4}+\frac{26.574}{(1+s)^3}-\frac{45.482}{(1+s)^2}+\frac{49.13}{1+s}+\frac{11.9964}{(2+s)^4}-\frac{5.26}{(2+s)^3}+\frac{45.482}{(2+s)^2}-\frac{79.9}{2+s}-\right.\nonumber\\
&&\frac{4.9518}{(3+s)^4}-\frac{21.314}{(3+s)^3}+\frac{26.463}{3+s}+\frac{3.7976}{4+s}+\frac{0.5094}{5+s}-\frac{9.517\left(\gamma _E+\psi (2+s)\right)}{1+s}+\nonumber\\
&&\frac{9.517 \left(\gamma _E+\psi (3+s)\right)}{2+s}+\frac{1.7805 \left(\pi ^2+6 \left(\gamma _E+\psi (2+s)\right){}^2-6 \psi '(2+s)\right)}{6+6
s}-\nonumber\\
&&\left.\frac{1.7805 \left(\pi ^2+6 \left(\gamma _E+\psi (3+s)\right){}^2-6 \psi '(3+s)\right)}{12+6 s}\right]+\nonumber\\
&&f \Bigg[-\frac{2752}{9 (1+s)^5}+\frac{545.5188}{(1+s)^4}-\frac{737.2}{(1+s)^3}+\frac{739}{(1+s)^2}-\frac{1362.6}{1+s}+\frac{2752}{9 (2+s)^5}-\frac{56.5188}{(2+s)^4}+\nonumber\\
&&\frac{1437}{(2+s)^3}-\frac{739}{(2+s)^2}+\frac{2980}{2+s}-\frac{489}{(3+s)^4}-\frac{699.8}{(3+s)^3}-\frac{2292.2}{3+s}+\frac{842.21}{4+s}-\frac{167.41}{5+s}+\nonumber\\
&&\frac{204.76 \left(\gamma _E+\psi (2+s)\right)}{1+s}-\frac{204.76\left(\gamma _E+\psi (3+s)\right)}{2+s}-\nonumber\\
&&\frac{232.57}{(1+s)^2 (2+s)^3}\left(5+\gamma _E (1+s)^2 (2+s)+2 s (3+s)+(1+s)^2 (2+s) \psi (3+s)-\right.\nonumber\\
&&\left.(1+s)^2 (2+s)^2 \psi '(1+s)\right)-\frac{12.61 \left(\pi ^2+6 \left(\gamma _E+\psi (2+s)\right){}^2-6 \psi '(2+s)\right)}{6+6 s}+\nonumber\\
&&\frac{232.57 \left(\gamma _E+\frac{1}{1+s}+\psi (1+s)-(1+s) \psi '(2+s)\right)}{(1+s)^2}+\nonumber\\
&&\frac{12.61 \left(\pi ^2+6 \left(\gamma _E+\psi (3+s)\right){}^2-6 \psi '(3+s)\right)}{12+6 s}+\nonumber\\
&&\frac{1}{1+s}3.2705\left(2 \gamma _E{}^3+\gamma _E \pi ^2+6 \gamma _E \psi (2+s)^2+2 \psi (2+s)^3+\psi (2+s) \left(6 \gamma _E{}^2+\pi ^2-6
\psi '(2+s)\right)-\right.\nonumber\\
&&\left.6 \gamma _E \psi '(2+s)+2 \psi ''(2+s)+4 \zeta (3)\right)-\nonumber\\
&&\frac{1}{2+s}3.2705 \left(2 \gamma _E{}^3+\gamma _E \pi ^2+6 \gamma _E \psi (3+s)^2+2 \psi (3+s)^3+\psi (3+s) \left(6 \gamma _E{}^2+\pi ^2-6
\psi '(3+s)\right)-\right.\nonumber\\
&&\left.6 \gamma _E \psi '(3+s)+2 \psi ''(3+s)+4 \zeta (3)\right)\Bigg].\nonumber
\end{eqnarray*}

\begin{eqnarray*}
&&\Delta \Phi _{NS}^{\rm NNLO}=\hspace{14 cm}\text{(A.2)}\\
&&1295.47\, +\frac{928}{27 (1+s)^5}-\frac{640}{3 (1+s)^4}+\frac{798.4}{(1+s)^3}-\frac{1465.2}{(1+s)^2}+\frac{1860.2}{1+s}-\frac{3505}{2+s}+\nonumber\\
&&\frac{297}{3+s}-\frac{433.2}{4+s}-1147.898\left(\gamma _E +\psi (1+s)\right)-\frac{714.1 \left(\gamma _E +\psi (2+s)\right)}{1+s}+\nonumber\\
&&\frac{684 \left(\gamma _E+\frac{1}{1+s}+\psi (1+s)-(1+s) \psi '(2+s)\right)}{(1+s)^2}-\nonumber\\
&&\frac{251.2}{(1+s)^4}\left(4+2 \gamma _E (1+s)+2 (1+s) \psi (1+s)-2 (1+s)^2 \psi '(1+s)+(1+s)^3 \psi ''(2+s)\right)+\nonumber\\
&&f \left[-173.933+\frac{512}{27 (1+s)^4}-\frac{2144}{27 (1+s)^3}+\frac{172.69}{(1+s)^2}-\frac{216.62}{1+s}+\frac{6.816}{(2+s)^4}+\frac{406.5}{2+s}+\right.\nonumber\\
&&\frac{77.89}{3+s}+\frac{34.76}{4+s}+183.187 \left(\gamma _E +\psi (1+s)\right)+\frac{5120 \left(\gamma _E +\psi (2+s)\right)}{81 (1+s)}-\nonumber\\
&&\left.\frac{65.43}{(1+s)^3}\left(1+\gamma _E+\gamma _E s+(1+s) \psi (2+s)-(1+s)^2 \psi '(1+s)\right)\right]+\nonumber\\
&&\frac{64}{81}f^2 \left[-\frac{51}{16}+\frac{5 \pi ^2}{6}+\frac{3}{2 (1+s)^3}-\frac{11}{2 (1+s)^2}+\frac{7}{1+s}-\frac{3}{2 (2+s)^3}+\frac{11}{2 (2+s)^2}-\frac{6}{2+s}+\right.\nonumber\\
&&\left.\left(\gamma _E +\psi (1+s)\right)-3 \zeta (3)-5 \psi '(2+s)-\frac{3}{2} \psi ''(2+s)\right].\nonumber
\end{eqnarray*}

\begin{eqnarray*}
&&\Delta \Theta _f^{\rm NNLO}=\hspace{14 cm}\text{(A.3)}\\
&&f \Bigg[-\frac{1208}{(1+s)^5}+\frac{2313.84}{(1+s)^4}-\frac{1789.6}{(1+s)^3}+\frac{1461.2}{(1+s)^2}-\frac{2972.4}{1+s}+\frac{439.8}{(2+s)^4}+\frac{2290.6}{(2+s)^3}+\frac{4672}{2+s}-\frac{1221.6}{3+s}-\nonumber\\
&&\frac{18}{4+s}-\frac{278.32 \left(\gamma _E+\psi (2+s)\right)}{1+s}-\frac{90.26 \left(\pi ^2+6 \left(\gamma _E+\psi (2+s)\right){}^2-6 \psi
'(2+s)\right)}{6+6 s}+\nonumber\\
&&\frac{825.4\left(\gamma _E+\frac{1}{1+s}+\psi (1+s)-(1+s) \psi '(2+s)\right)}{(1+s)^2}+\nonumber\\
&&\frac{1}{1+s}2.65 \left(2 \gamma _E{}^3+\gamma _E \pi ^2+6 \gamma _E \psi (2+s)^2+2 \psi (2+s)^3+\psi (2+s) \left(6 \gamma _E{}^2+\pi ^2-6 \psi
'(2+s)\right)-\right.\nonumber\\
&&\left.6 \gamma _E \psi '(2+s)+2 \psi ''(2+s)+4 \zeta (3)\right)+\nonumber\\
&&\frac{1}{1+s}0.1892\left(20 \gamma _E{}^4+20 \gamma _E{}^2 \pi ^2+3 \pi ^4+80 \gamma _E \psi (2+s)^3+20 \psi (2+s)^4+20 \psi (2+s)^2 \left(6
\gamma _E{}^2+\pi ^2-6 \psi '(2+s)\right)-\right.\nonumber\\
&&20 \left(6 \gamma _E{}^2+\pi ^2\right) \psi '(2+s)+60 (\psi '(2+s))^2+80 \gamma _E \psi ''(2+s)-20  \psi '''(2+s)+\nonumber\\
&&\left.160 \gamma _E \zeta (3)+40 \psi (2+s) \left(2 \gamma _E{}^3+\gamma _E \pi ^2-6 \gamma _E \psi '(2+s)+2 \psi ''(2+s)+4 \zeta
(3)\right)\right)\Bigg]+\nonumber\\
&&f^2 \Bigg[\frac{128}{3 (1+s)^5}-\frac{184.434}{(1+s)^4}+\frac{393.92}{(1+s)^3}-\frac{526.3}{(1+s)^2}+\frac{499.65}{1+s}-\frac{61.116}{(2+s)^4}+\frac{358.2}{(2+s)^3}-\frac{432.18}{2+s}-\frac{141.63}{3+s}-\nonumber\\
&&\frac{11.34}{4+s}+\frac{6.256 \left(\gamma _E+\psi (2+s)\right)}{1+s}-\frac{1}{(1+s)^3}47.3 \left(1+\gamma _E+\gamma _E s+(1+s) \psi (2+s)-(1+s)^2
\psi '(1+s)\right)+\nonumber\\
&&\frac{7.32 \left(\pi ^2+6 \left(\gamma _E+\psi (2+s)\right){}^2-6 \psi '(2+s)\right)}{6+6 s}-\nonumber\\
&&\frac{1}{1+s}0.3687 \left(2 \gamma _E{}^3+\gamma _E \pi ^2+6 \gamma _E \psi (2+s)^2+2 \psi (2+s)^3+\psi (2+s) \left(6 \gamma _E{}^2+\pi ^2-6
\psi '(2+s)\right)-\right.\nonumber\\
&&\left.6 \gamma _E \psi '(2+s)+2 \psi ''(2+s)+4 \zeta (3)\right)\Bigg].\nonumber
\end{eqnarray*}

\begin{eqnarray*}
&&\Delta \Theta _g^{\rm NNLO}=\hspace{14 cm}\text{(A.4)}\\
&&\frac{92096}{27 (1+s)^5}-\frac{5328.018}{(1+s)^4}+\frac{4280}{(1+s)^3}-\frac{4046.6}{(1+s)^2}+\frac{6159}{1+s}-\frac{1050.6}{(2+s)^4}-\frac{1701.4}{(2+s)^3}-\frac{3825.9}{2+s}+\frac{1942}{3+s}-\nonumber\\
&&\frac{742.1}{4+s}-\frac{1843.7 \left(\gamma _E+\psi (2+s)\right)}{1+s}+\frac{451.55 \left(\pi ^2+6 \left(\gamma _E+\psi (2+s)\right){}^2-6 \psi
'(2+s)\right)}{6+6 s}-\nonumber\\
&&\frac{1424.8 \left(\gamma _E+\frac{1}{1+s}+\psi (1+s)-(1+s) \psi '(2+s)\right)}{(1+s)^2}-\nonumber\\
&&\frac{1}{1+s}29.65 \left(2 \gamma _E{}^3+\gamma _E \pi ^2+6 \gamma _E \psi (2+s)^2+2 \psi (2+s)^3+\psi (2+s) \left(6 \gamma _E{}^2+\pi ^2-6
\psi '(2+s)\right)-\right.\nonumber\\
&&\left.6 \gamma _E \psi '(2+s)+2 \psi ''(2+s)+4 \zeta (3)\right)+\nonumber\\
&&\frac{1}{1+s}0.25715\left(20 \gamma _E{}^4+20 \gamma _E{}^2 \pi ^2+3 \pi ^4+80 \gamma _E \psi (2+s)^3+20 \psi (2+s)^4+20 \psi (2+s)^2 \left(6
\gamma _E{}^2+\pi ^2-6 \psi '(2+s)\right)-\right.\nonumber\\
&&20 \left(6 \gamma _E{}^2+\pi ^2\right) \psi '(2+s)+60 (\psi '(2+s))^2+80 \gamma _E \psi ''(2+s)-20 \psi '''(2+s)+160 \gamma _E
\zeta (3)+\nonumber\\
&&\left.40 \psi (2+s) \left(2 \gamma _E{}^3+\gamma _E \pi ^2-6 \gamma _E \psi '(2+s)+2 \psi ''(2+s)+4 \zeta (3)\right)\right)+\nonumber\\
&&f \Bigg[-\frac{1024}{9 (1+s)^5}+\frac{236.3232}{(1+s)^4}-\frac{404.92}{(1+s)^3}+\frac{308.98}{(1+s)^2}-\frac{301.07}{1+s}+\frac{180.138}{(2+s)^4}-\frac{253.06}{(2+s)^3}-\frac{296}{2+s}+\frac{406.13}{3+s}-\nonumber\\
&&\frac{101.62}{4+s}+\frac{171.78\left(\gamma _E+\psi (2+s)\right)}{1+s}-\frac{16.18 \left(1+\gamma _E+\gamma _E s+(1+s) \psi (2+s)-(1+s)^2 \psi
'(1+s)\right)}{(1+s)^3}-\nonumber\\
&&\frac{47.86 \left(\pi ^2+6 \left(\gamma _E+\psi (2+s)\right){}^2-6 \psi '(2+s)\right)}{6+6 s}+\nonumber\\
&&\frac{1}{1+s}2.4815\left(2 \gamma _E{}^3+\gamma _E \pi ^2+6 \gamma _E \psi (2+s)^2+2 \psi (2+s)^3+\psi (2+s) \left(6 \gamma _E{}^2+\pi ^2-6
\psi '(2+s)\right)-\right.\nonumber\\
&&\left.6 \gamma _E \psi '(2+s)+2 \psi ''(2+s)+4 \zeta (3)\right)\Bigg]+\nonumber\\
&&\frac{16}{27}f^2 \left[-\frac{12}{1+s}+\frac{10}{2+s}-\frac{8\left(\gamma _E+\psi (2+s)\right)}{1+s}-\frac{2 \left(\gamma _E+\psi (3+s)\right)}{2+s}+\frac{6
\left(\pi ^2+6 \left(\gamma _E+\psi (2+s)\right){}^2-6 \psi '(2+s)\right)}{6+6 s}-\right.\nonumber\\
&&\left.\frac{3 \left(\pi ^2+6 \left(\gamma _E+\psi (3+s)\right){}^2-6 \psi '(3+s)\right)}{12+6 s}\right].\nonumber
\end{eqnarray*}

\begin{eqnarray*}
&&\Delta \Phi _g^{\rm NNLO}=\hspace{14 cm}\text{(A.5)}\\
&&4427.762+\frac{12096}{(1+s)^5}-\frac{22665}{(1+s)^4}+\frac{21804}{(1+s)^3}-\frac{23091}{(1+s)^2}+\frac{30988}{1+s}-\frac{7002}{(2+s)^4}-\frac{1726}{(2+s)^3}-\frac{39925}{2+s}+\nonumber\\
&&\frac{13447}{3+s}-\frac{4576}{4+s}-2643.521\left(\gamma _E+\psi (1+s)\right)+\frac{9446\left(\gamma _E+\psi (2+s)\right)}{1+s}-\frac{13247 \left(\gamma
_E+\psi (3+s)\right)}{2+s}-\nonumber\\
&&\frac{12292 \left(1+\gamma _E+\gamma _E s+(1+s) \psi (2+s)-(1+s)^2 \psi '(1+s)\right)}{(1+s)^3}+\nonumber\\
&&f \left[-528.536-\frac{6128}{9 (1+s)^5}+\frac{2146.788}{(1+s)^4}-\frac{3754.4}{(1+s)^3}+\frac{3524}{(1+s)^2}-\frac{1173.5}{1+s}-\frac{786}{(2+s)^4}+\frac{1226.2}{(2+s)^3}+\frac{2648.6}{2+s}-\right.\nonumber\\
&&\frac{2160.8}{3+s}+\frac{1251.7}{4+s}+412.172\left(\gamma _E+\psi (1+s)\right)+\frac{7041.7\left(\gamma _E+\psi (2+s)\right)}{1+s}-\frac{6746
\left(\gamma _E+\psi (3+s)\right)}{2+s}-\nonumber\\
&&\left.\frac{7932 \left(1+\gamma _E+\gamma _E s+(1+s) \psi (2+s)-(1+s)^2 \psi '(1+s)\right)}{(1+s)^3}\right]+\nonumber\\
&&f^2 \left[6.4607\, +\frac{7.0854}{(1+s)^4}-\frac{13.358}{(1+s)^3}+\frac{13.29}{(1+s)^2}-\frac{16.606}{1+s}+\frac{31.528}{(2+s)^3}+\frac{32.905}{2+s}-\frac{18.3}{3+s}+\frac{2.637}{4+s}+\right.\nonumber\\
&&\left.\frac{16 \left(\gamma _E+\psi (1+s)\right)}{9}+\frac{0.21 \left(\gamma _E+\psi (2+s)\right)}{1+s}-\frac{16.944 \left(1+\gamma _E+\gamma
_E s+(1+s) \psi (2+s)-(1+s)^2 \psi '(1+s)\right)}{(1+s)^3}\right]\,. \nonumber
\end{eqnarray*}

The first moment of $\Delta \Phi_g^{\text {NNLO}} (s=0, f=3)$ is $\beta_2^{\overline{\text {MS}}} = 643.833$

\section{ The coefficient functions of singlet distributions in the Laplace $s$-space at the NNLO approximation }\label{AppendixB}

Here, we present the coefficient functions of singlet distributions, $k_{ff}$, $k_{fg}$, $k_{gf}$ and $k_{gg}$ in the Laplace $s$-space at the NNLO approximation. They are given by:

\begin{eqnarray*}
&&k_{ff}=\hspace{14 cm}\text{(B.1)}\\
&&\Bigg(e^{\frac{1}{2} \left(-4 b_1+\Delta \Phi _f+\Delta \Phi _g-R\right) \tau } \Big[8 b_1{}^5 e^{2 b_1 \tau } \left(-\Delta \Phi _f+\Delta
\Phi _g+R+e^{R \tau } \left(\Delta \Phi _f-\Delta \Phi _g+R\right)\right)+\\
&&4 a_1 b_1{}^4 \left(e^{\left(2 b_1+R\right) \tau } \left(2 \Delta \Phi _f^{\text{nlo}}+\left(4 a_0+a_1\right) \Delta \Phi _f^{\text{nnlo}}\right)
\left(\Delta \Phi _f-\Delta \Phi _g+R\right)+\right.\\
&&e^{2 b_1 \tau } \left(2 \Delta \Phi _f^{\text{nlo}}+\left(4 a_0+a_1\right) \Delta \Phi _f^{\text{nnlo}}\right) \left(-\Delta \Phi _f+\Delta
\Phi _g+R\right)-\\
&&a_1 e^{R \tau } \left(\Delta \Phi _f^{\text{nnlo}} \left(\Delta \Phi _f-\Delta \Phi _g+R\right)+2 \Delta \Theta _f^{\text{nnlo}} \Delta \Theta
_g\right)+\\
&&a_1 \left(\Delta \Phi _f \Delta \Phi _f^{\text{nnlo}}-\Delta \Phi _f^{\text{nnlo}} \left(\Delta \Phi _g+R\right)+2 \Delta \Theta _f^{\text{nnlo}}
\Delta \Theta _g\right)-\\
&&2 e^{b_1 \tau } \left(-\left(\Delta \Phi _f^{\text{nlo}}+2 a_0 \Delta \Phi _f^{\text{nnlo}}\right) \left(\Delta \Phi _f-\Delta \Phi _g-R\right)-2
\left(\Delta \Theta _f^{\text{nlo}}+2 a_0 \Delta \Theta _f^{\text{nnlo}}\right) \Delta \Theta _g\right)-\\
&&\left.2 e^{\left(b_1+R\right) \tau } \left(\left(\Delta \Phi _f^{\text{nlo}}+2 a_0 \Delta \Phi _f^{\text{nnlo}}\right) \left(\Delta \Phi _f-\Delta
\Phi _g+R\right)+2 \left(\Delta \Theta _f^{\text{nlo}}+2 a_0 \Delta \Theta _f^{\text{nnlo}}\right) \Delta \Theta _g\right)\right)-\\
&&2 b_1{}^3 \left(a_1{}^2 \left(-2 \Delta \Phi _f^{\text{nnlo}} \Delta \Theta _f+\left(\Delta \Phi _f-\Delta \Phi _g+R\right) \Delta \Theta _f^{\text{nnlo}}\right)
\Delta \Theta _g-\right.\\
&&4 a_1 e^{b_1 \tau } \left(2 \Delta \Phi _f^{\text{nlo}} \Delta \Theta _f+4 a_0 \Delta \Phi _f^{\text{nnlo}} \Delta \Theta _f-\left(\Delta \Phi
_f-\Delta \Phi _g+R\right) \Delta \Theta _f^{\text{nlo}}-2 a_0 \left(\Delta \Phi _f-\Delta \Phi _g+R\right) \Delta \Theta _f^{\text{nnlo}}\right)
\\
&&\Delta \Theta _g+a_1{}^2 e^{R \tau } \left(2 \Delta \Phi _f^{\text{nnlo}} \Delta \Theta _f+\left(-\Delta \Phi _f+\Delta \Phi _g+R\right) \Delta
\Theta _f^{\text{nnlo}}\right) \Delta \Theta _g+\\
&&4 a_1 e^{\left(b_1+R\right) \tau } \left(2 \Delta \Phi _f^{\text{nlo}} \Delta \Theta _f+\left(-\Delta \Phi _f+\Delta \Phi _g+R\right) \Delta
\Theta _f^{\text{nlo}}+\right.\\
&&\left.2 a_0 \left(2 \Delta \Phi _f^{\text{nnlo}} \Delta \Theta _f+\left(-\Delta \Phi _f+\Delta \Phi _g+R\right) \Delta \Theta _f^{\text{nnlo}}\right)\right)
\Delta \Theta _g-\\
&&e^{2 b_1 \tau } \left(5 \left(\Delta \Phi _f-\Delta \Phi _g\right){}^2 \left(\Delta \Phi _f-\Delta \Phi _g-R\right)+\right.\\
&&\left(2 \left(10 \Delta \Phi _f+a_1 \left(4 \Delta \Phi _f^{\text{nlo}}+8 a_0 \Delta \Phi _f^{\text{nnlo}}+a_1 \Delta \Phi _f^{\text{nnlo}}\right)-10
\left(\Delta \Phi _g+R\right)\right) \Delta \Theta _f+\right.\\
&&\left.\left.4 a_1 \left(-\Delta \Phi _f+\Delta \Phi _g+R\right) \Delta \Theta _f^{\text{nlo}}-a_1 \left(8 a_0+a_1\right) \left(\Delta \Phi _f-\Delta
\Phi _g-R\right) \Delta \Theta _f^{\text{nnlo}}\right) \Delta \Theta _g\right)+\\
&&e^{\left(2 b_1+R\right) \tau } \left(5 \left(\Delta \Phi _f-\Delta \Phi _g\right){}^2 \left(\Delta \Phi _f-\Delta \Phi _g+R\right)+\right.\\
&&\left(2 \left(10 \Delta \Phi _f+a_1 \left(4 \Delta \Phi _f^{\text{nlo}}+8 a_0 \Delta \Phi _f^{\text{nnlo}}+a_1 \Delta \Phi _f^{\text{nnlo}}\right)+10
\left(-\Delta \Phi _g+R\right)\right) \Delta \Theta _f-\right.\\
&&\left.\left.\left.4 a_1 \left(\Delta \Phi _f-\Delta \Phi _g+R\right) \Delta \Theta _f^{\text{nlo}}-a_1 \left(8 a_0+a_1\right) \left(\Delta \Phi
_f-\Delta \Phi _g+R\right) \Delta \Theta _f^{\text{nnlo}}\right) \Delta \Theta _g\right)\right)+\\
&&2 b_1 \left(\left(\Delta \Phi _f-\Delta \Phi _g\right){}^2+4 \Delta \Theta _f \Delta \Theta _g\right)\allowdisplaybreaks[1] \\
&&\left(a_1{}^2 \left(-2 \Delta \Phi _f^{\text{nnlo}} \Delta \Theta _f+\left(\Delta \Phi _f-\Delta \Phi _g+R\right) \Delta \Theta _f^{\text{nnlo}}\right)
\Delta \Theta _g-\right.\\
&&a_1 e^{b_1 \tau } \left(2 \Delta \Phi _f^{\text{nlo}} \Delta \Theta _f+4 a_0 \Delta \Phi _f^{\text{nnlo}} \Delta \Theta _f-\left(\Delta \Phi
_f-\Delta \Phi _g+R\right) \Delta \Theta _f^{\text{nlo}}-2 a_0 \left(\Delta \Phi _f-\Delta \Phi _g+R\right) \Delta \Theta _f^{\text{nnlo}}\right)
\Delta \Theta _g+\\
&&a_1{}^2 e^{R \tau } \left(2 \Delta \Phi _f^{\text{nnlo}} \Delta \Theta _f+\left(-\Delta \Phi _f+\Delta \Phi _g+R\right) \Delta \Theta _f^{\text{nnlo}}\right)
\Delta \Theta _g+\\
&&a_1 e^{\left(b_1+R\right) \tau } \left(2 \Delta \Phi _f^{\text{nlo}} \Delta \Theta _f+\left(-\Delta \Phi _f+\Delta \Phi _g+R\right) \Delta \Theta
_f^{\text{nlo}}+\right.\\
&&\left.2 a_0 \left(2 \Delta \Phi _f^{\text{nnlo}} \Delta \Theta _f+\left(-\Delta \Phi _f+\Delta \Phi _g+R\right) \Delta \Theta _f^{\text{nnlo}}\right)\right)
\Delta \Theta _g+\\
&&e^{\left(2 b_1+R\right) \tau } \left(\left(\Delta \Phi _f-\Delta \Phi _g\right){}^2 \left(\Delta \Phi _f-\Delta \Phi _g+R\right)+\right.\\
&&\left(4 \Delta \Phi _f \Delta \Theta _f+2 a_1 \Delta \Phi _f^{\text{nlo}} \Delta \Theta _f+4 a_0 a_1 \Delta \Phi _f^{\text{nnlo}} \Delta \Theta
_f+2 a_1{}^2 \Delta \Phi _f^{\text{nnlo}} \Delta \Theta _f-4 \Delta \Phi _g \Delta \Theta _f+\right.\\
&&\left.4 R \Delta \Theta _f-a_1 \Delta \Phi _f \Delta \Theta _f^{\text{nlo}}+a_1 \Delta \Phi _g \Delta \Theta _f^{\text{nlo}}-a_1 R \Delta \Theta
_f^{\text{nlo}}-a_1 \left(2 a_0+a_1\right) \left(\Delta \Phi _f-\Delta \Phi _g+R\right) \Delta \Theta _f^{\text{nnlo}}\right) \\
&&\left.\Delta \Theta _g\right)-\\
&&e^{2 b_1 \tau } \left(\left(\Delta \Phi _f-\Delta \Phi _g\right){}^2 \left(\Delta \Phi _f-\Delta \Phi _g-R\right)+\right.\\
&&\left(4 \Delta \Phi _f \Delta \Theta _f+2 a_1 \Delta \Phi _f^{\text{nlo}} \Delta \Theta _f+4 a_0 a_1 \Delta \Phi _f^{\text{nnlo}} \Delta \Theta
_f+2 a_1{}^2 \Delta \Phi _f^{\text{nnlo}} \Delta \Theta _f-4 \Delta \Phi _g \Delta \Theta _f-\right.\\
&&\left.4 R \Delta \Theta _f-a_1 \Delta \Phi _f \Delta \Theta _f^{\text{nlo}}+a_1 \Delta \Phi _g \Delta \Theta _f^{\text{nlo}}+a_1 R \Delta \Theta
_f^{\text{nlo}}+a_1 \left(2 a_0+a_1\right) \left(-\Delta \Phi _f+\Delta \Phi _g+R\right) \Delta \Theta _f^{\text{nnlo}}\right) \\
&&\left.\left.\Delta \Theta _g\right)\right)+a_1 \left(\left(\Delta \Phi _f-\Delta \Phi _g\right){}^2+4 \Delta \Theta _f \Delta \Theta _g\right)
\\
&&\left(a_1 \left(\Delta \Phi _f^{\text{nnlo}} \left(\Delta \Phi _f-\Delta \Phi _g\right){}^2 \left(\Delta \Phi _f-\Delta \Phi _g-R\right)+\right.\right.\\
&&\left(2 \Delta \Phi _f^{\text{nnlo}} \left(2 \Delta \Phi _f-2 \Delta \Phi _g-R\right) \Delta \Theta _f+\left(\Delta \Phi _f-\Delta \Phi _g\right)
\left(\Delta \Phi _f-\Delta \Phi _g-R\right) \Delta \Theta _f^{\text{nnlo}}\right) \Delta \Theta _g+\\
&&\left.4 \Delta \Theta _f \Delta \Theta _f^{\text{nnlo}} \Delta \Theta _g{}^2\right)-\\
&&a_1 e^{R \tau } \left(\Delta \Phi _f^{\text{nnlo}} \left(\Delta \Phi _f-\Delta \Phi _g\right){}^2 \left(\Delta \Phi _f-\Delta \Phi _g+R\right)+\right.\\
&&\left(2 \Delta \Phi _f^{\text{nnlo}} \left(2 \Delta \Phi _f-2 \Delta \Phi _g+R\right) \Delta \Theta _f+\left(\Delta \Phi _f-\Delta \Phi _g\right)
\left(\Delta \Phi _f-\Delta \Phi _g+R\right) \Delta \Theta _f^{\text{nnlo}}\right) \Delta \Theta _g+\\
&&\left.4 \Delta \Theta _f \Delta \Theta _f^{\text{nnlo}} \Delta \Theta _g{}^2\right)-\\
&&2 e^{\left(b_1+R\right) \tau } \left(\left(\Delta \Phi _f^{\text{nlo}}+2 a_0 \Delta \Phi _f^{\text{nnlo}}\right) \left(\Delta \Phi _f-\Delta
\Phi _g\right){}^2 \left(\Delta \Phi _f-\Delta \Phi _g+R\right)+\right.\\
&&\left(4 \Delta \Phi _f \Delta \Phi _f^{\text{nlo}} \Delta \Theta _f+8 a_0 \Delta \Phi _f \Delta \Phi _f^{\text{nnlo}} \Delta \Theta _f-4 \Delta
\Phi _f^{\text{nlo}} \Delta \Phi _g \Delta \Theta _f-8 a_0 \Delta \Phi _f^{\text{nnlo}} \Delta \Phi _g \Delta \Theta _f+\right.\\
&&2 \Delta \Phi _f^{\text{nlo}} R \Delta \Theta _f+4 a_0 \Delta \Phi _f^{\text{nnlo}} R \Delta \Theta _f+\Delta \Phi _f{}^2 \Delta \Theta _f^{\text{nlo}}-2
\Delta \Phi _f \Delta \Phi _g \Delta \Theta _f^{\text{nlo}}+\Delta \Phi _g{}^2 \Delta \Theta _f^{\text{nlo}}+\\
&&\left.\Delta \Phi _f R \Delta \Theta _f^{\text{nlo}}-\Delta \Phi _g R \Delta \Theta _f^{\text{nlo}}+2 a_0 \left(\Delta \Phi _f-\Delta \Phi _g\right)
\left(\Delta \Phi _f-\Delta \Phi _g+R\right) \Delta \Theta _f^{\text{nnlo}}\right) \Delta \Theta _g+\\
&&\left.4 \Delta \Theta _f \left(\Delta \Theta _f^{\text{nlo}}+2 a_0 \Delta \Theta _f^{\text{nnlo}}\right) \Delta \Theta _g{}^2\right)+\\
&&2 e^{b_1 \tau } \left(\left(\Delta \Phi _f^{\text{nlo}}+2 a_0 \Delta \Phi _f^{\text{nnlo}}\right) \left(\Delta \Phi _f-\Delta \Phi _g\right){}^2
\left(\Delta \Phi _f-\Delta \Phi _g-R\right)+\right.\\
&&\left(4 \Delta \Phi _f \Delta \Phi _f^{\text{nlo}} \Delta \Theta _f+8 a_0 \Delta \Phi _f \Delta \Phi _f^{\text{nnlo}} \Delta \Theta _f-4 \Delta
\Phi _f^{\text{nlo}} \Delta \Phi _g \Delta \Theta _f-8 a_0 \Delta \Phi _f^{\text{nnlo}} \Delta \Phi _g \Delta \Theta _f-\right.\\
&&2 \Delta \Phi _f^{\text{nlo}} R \Delta \Theta _f-4 a_0 \Delta \Phi _f^{\text{nnlo}} R \Delta \Theta _f+\Delta \Phi _f{}^2 \Delta \Theta _f^{\text{nlo}}-2
\Delta \Phi _f \Delta \Phi _g \Delta \Theta _f^{\text{nlo}}+\Delta \Phi _g{}^2 \Delta \Theta _f^{\text{nlo}}-\\
&&\left.\Delta \Phi _f R \Delta \Theta _f^{\text{nlo}}+\Delta \Phi _g R \Delta \Theta _f^{\text{nlo}}+2 a_0 \left(-\Delta \Phi _f+\Delta \Phi
_g\right) \left(-\Delta \Phi _f+\Delta \Phi _g+R\right) \Delta \Theta _f^{\text{nnlo}}\right) \Delta \Theta _g+\\
&&\left.4 \Delta \Theta _f \left(\Delta \Theta _f^{\text{nlo}}+2 a_0 \Delta \Theta _f^{\text{nnlo}}\right) \Delta \Theta _g{}^2\right)+\\
&&e^{2 b_1 \tau } \left(-\left(2 \Delta \Phi _f^{\text{nlo}}+\left(4 a_0+a_1\right) \Delta \Phi _f^{\text{nnlo}}\right) \left(\Delta \Phi _f-\Delta
\Phi _g\right){}^2 \left(\Delta \Phi _f-\Delta \Phi _g-R\right)-\right.\\
&&\left(2 \left(2 \Delta \Phi _f^{\text{nlo}}+\left(4 a_0+a_1\right) \Delta \Phi _f^{\text{nnlo}}\right) \left(2 \Delta \Phi _f-2 \Delta \Phi
_g-R\right) \Delta \Theta _f+\right.\\
&&\left.2 \left(\Delta \Phi _f-\Delta \Phi _g\right) \left(\Delta \Phi _f-\Delta \Phi _g-R\right) \Delta \Theta _f^{\text{nlo}}+\left(4 a_0+a_1\right)
\left(\Delta \Phi _f-\Delta \Phi _g\right) \left(\Delta \Phi _f-\Delta \Phi _g-R\right) \Delta \Theta _f^{\text{nnlo}}\right) \\
&&\left.\Delta \Theta _g-4 \Delta \Theta _f \left(2 \Delta \Theta _f^{\text{nlo}}+\left(4 a_0+a_1\right) \Delta \Theta _f^{\text{nnlo}}\right)
\Delta \Theta _g{}^2\right)+\\
&&e^{\left(2 b_1+R\right) \tau } \left(\left(2 \Delta \Phi _f^{\text{nlo}}+\left(4 a_0+a_1\right) \Delta \Phi _f^{\text{nnlo}}\right) \left(\Delta
\Phi _f-\Delta \Phi _g\right){}^2 \left(\Delta \Phi _f-\Delta \Phi _g+R\right)+\right.\\
&&\left(2 \left(2 \Delta \Phi _f^{\text{nlo}}+\left(4 a_0+a_1\right) \Delta \Phi _f^{\text{nnlo}}\right) \left(2 \Delta \Phi _f-2 \Delta \Phi
_g+R\right) \Delta \Theta _f+\right.\\
&&\left.2 \left(\Delta \Phi _f-\Delta \Phi _g\right) \left(\Delta \Phi _f-\Delta \Phi _g+R\right) \Delta \Theta _f^{\text{nlo}}+\left(4 a_0+a_1\right)
\left(\Delta \Phi _f-\Delta \Phi _g\right) \left(\Delta \Phi _f-\Delta \Phi _g+R\right) \Delta \Theta _f^{\text{nnlo}}\right) \\
&&\left.\left.\Delta \Theta _g+4 \Delta \Theta _f \left(2 \Delta \Theta _f^{\text{nlo}}+\left(4 a_0+a_1\right) \Delta \Theta _f^{\text{nnlo}}\right)
\Delta \Theta _g{}^2\right)\right)+\\
&&a_1 b_1{}^2 \\
&&\Big\{a_1 \left(-5 \Delta \Phi _f^{\text{nnlo}} \left(\Delta \Phi _f-\Delta \Phi _g\right){}^2 \left(\Delta \Phi _f-\Delta \Phi _g-R\right)+\right.\\
&&\left(2 \Delta \Phi _f^{\text{nnlo}} \left(-10 \Delta \Phi _f+10 \Delta \Phi _g+9 R\right) \Delta \Theta _f-\left(\Delta \Phi _f-\Delta \Phi
_g\right) \left(9 \Delta \Phi _f-9 \Delta \Phi _g-R\right) \Delta \Theta _f^{\text{nnlo}}\right) \Delta \Theta _g-\allowdisplaybreaks[1]\\
&&\left.36 \Delta \Theta _f \Delta \Theta _f^{\text{nnlo}} \Delta \Theta _g{}^2\right)+\\
&&a_1 e^{R \tau } \left(5 \Delta \Phi _f^{\text{nnlo}} \left(\Delta \Phi _f-\Delta \Phi _g\right){}^2 \left(\Delta \Phi _f-\Delta \Phi _g+R\right)+\right.\\
&&\left(2 \Delta \Phi _f^{\text{nnlo}} \left(10 \Delta \Phi _f-10 \Delta \Phi _g+9 R\right) \Delta \Theta _f+\left(\Delta \Phi _f-\Delta \Phi
_g\right) \left(9 \Delta \Phi _f-9 \Delta \Phi _g+R\right) \Delta \Theta _f^{\text{nnlo}}\right) \Delta \Theta _g+\\
&&\left.36 \Delta \Theta _f \Delta \Theta _f^{\text{nnlo}} \Delta \Theta _g{}^2\right)+\\
&&2 e^{b_1 \tau } \left(-5 \left(\Delta \Phi _f^{\text{nlo}}+2 a_0 \Delta \Phi _f^{\text{nnlo}}\right) \left(\Delta \Phi _f-\Delta \Phi _g\right){}^2
\left(\Delta \Phi _f-\Delta \Phi _g-R\right)-\right.\\
&&2 \left(10 \Delta \Phi _f \Delta \Phi _f^{\text{nlo}} \Delta \Theta _f+20 a_0 \Delta \Phi _f \Delta \Phi _f^{\text{nnlo}} \Delta \Theta _f-10
\Delta \Phi _f^{\text{nlo}} \Delta \Phi _g \Delta \Theta _f-20 a_0 \Delta \Phi _f^{\text{nnlo}} \Delta \Phi _g \Delta \Theta _f-\right.\\
&&6 \Delta \Phi _f^{\text{nlo}} R \Delta \Theta _f-12 a_0 \Delta \Phi _f^{\text{nnlo}} R \Delta \Theta _f+3 \Delta \Phi _f{}^2 \Delta \Theta _f^{\text{nlo}}-6
\Delta \Phi _f \Delta \Phi _g \Delta \Theta _f^{\text{nlo}}+3 \Delta \Phi _g{}^2 \Delta \Theta _f^{\text{nlo}}-\\
&&\left.2 \Delta \Phi _f R \Delta \Theta _f^{\text{nlo}}+2 \Delta \Phi _g R \Delta \Theta _f^{\text{nlo}}+2 a_0 \left(\Delta \Phi _f-\Delta \Phi
_g\right) \left(3 \Delta \Phi _f-3 \Delta \Phi _g-2 R\right) \Delta \Theta _f^{\text{nnlo}}\right) \Delta \Theta _g-\\
&&\left.24 \Delta \Theta _f \left(\Delta \Theta _f^{\text{nlo}}+2 a_0 \Delta \Theta _f^{\text{nnlo}}\right) \Delta \Theta _g{}^2\right)+\\
&&2 e^{\left(b_1+R\right) \tau } \left(5 \left(\Delta \Phi _f^{\text{nlo}}+2 a_0 \Delta \Phi _f^{\text{nnlo}}\right) \left(\Delta \Phi _f-\Delta
\Phi _g\right){}^2 \left(\Delta \Phi _f-\Delta \Phi _g+R\right)+\right.\\
&&2 \left(10 \Delta \Phi _f \Delta \Phi _f^{\text{nlo}} \Delta \Theta _f+20 a_0 \Delta \Phi _f \Delta \Phi _f^{\text{nnlo}} \Delta \Theta _f-10
\Delta \Phi _f^{\text{nlo}} \Delta \Phi _g \Delta \Theta _f-20 a_0 \Delta \Phi _f^{\text{nnlo}} \Delta \Phi _g \Delta \Theta _f+\right.\\
&&6 \Delta \Phi _f^{\text{nlo}} R \Delta \Theta _f+12 a_0 \Delta \Phi _f^{\text{nnlo}} R \Delta \Theta _f+3 \Delta \Phi _f{}^2 \Delta \Theta _f^{\text{nlo}}-6
\Delta \Phi _f \Delta \Phi _g \Delta \Theta _f^{\text{nlo}}+3 \Delta \Phi _g{}^2 \Delta \Theta _f^{\text{nlo}}+\\
&&\left.2 \Delta \Phi _f R \Delta \Theta _f^{\text{nlo}}-2 \Delta \Phi _g R \Delta \Theta _f^{\text{nlo}}+2 a_0 \left(\Delta \Phi _f-\Delta \Phi
_g\right) \left(3 \Delta \Phi _f-3 \Delta \Phi _g+2 R\right) \Delta \Theta _f^{\text{nnlo}}\right) \Delta \Theta _g+\\
&&\left.24 \Delta \Theta _f \left(\Delta \Theta _f^{\text{nlo}}+2 a_0 \Delta \Theta _f^{\text{nnlo}}\right) \Delta \Theta _g{}^2\right)+\\
&&e^{2 b_1 \tau } \left(5 \left(2 \Delta \Phi _f^{\text{nlo}}+\left(4 a_0+a_1\right) \Delta \Phi _f^{\text{nnlo}}\right) \left(\Delta \Phi _f-\Delta
\Phi _g\right){}^2 \left(\Delta \Phi _f-\Delta \Phi _g-R\right)+\right.\\
&&\left(2 \left(10 \left(2 \Delta \Phi _f^{\text{nlo}}+\left(4 a_0+a_1\right) \Delta \Phi _f^{\text{nnlo}}\right) \left(\Delta \Phi _f-\Delta
\Phi _g\right)-3 \left(4 \Delta \Phi _f^{\text{nlo}}+8 a_0 \Delta \Phi _f^{\text{nnlo}}+3 a_1 \Delta \Phi _f^{\text{nnlo}}\right) R\right) \right.\\
&&\Delta \Theta _f+8 \left(\Delta \Phi _f-\Delta \Phi _g\right) \left(\Delta \Phi _f-\Delta \Phi _g-R\right) \Delta \Theta _f^{\text{nlo}}+\left(16
a_0+a_1\right) \left(\Delta \Phi _f-\Delta \Phi _g\right) \\
&&\left.\left.\left(\Delta \Phi _f-\Delta \Phi _g-R\right) \Delta \Theta _f^{\text{nnlo}}\right) \Delta \Theta _g+4 \Delta \Theta _f \left(8 \Delta
\Theta _f^{\text{nlo}}+\left(16 a_0+a_1\right) \Delta \Theta _f^{\text{nnlo}}\right) \Delta \Theta _g{}^2\right)-\\
&&e^{\left(2 b_1+R\right) \tau } \Big(5 \left(2 \Delta \Phi _f^{\text{nlo}}+\left(4 a_0+a_1\right) \Delta \Phi _f^{\text{nnlo}}\right) \left(\Delta
\Phi _f-\Delta \Phi _g\right){}^2 \left(\Delta \Phi _f-\Delta \Phi _g+R\right)+\\
&&\left(2 \left(10 \left(2 \Delta \Phi _f^{\text{nlo}}+\left(4 a_0+a_1\right) \Delta \Phi _f^{\text{nnlo}}\right) \left(\Delta \Phi _f-\Delta
\Phi _g\right)+3 \left(4 \Delta \Phi _f^{\text{nlo}}+8 a_0 \Delta \Phi _f^{\text{nnlo}}+3 a_1 \Delta \Phi _f^{\text{nnlo}}\right) R\right) \right.\\
&&\Delta \Theta _f+8 \left(\Delta \Phi _f-\Delta \Phi _g\right) \left(\Delta \Phi _f-\Delta \Phi _g+R\right) \Delta \Theta _f^{\text{nlo}}+\left(16
a_0+a_1\right) \left(\Delta \Phi _f-\Delta \Phi _g\right) \\
&&\left.\left.\left(\Delta \Phi _f-\Delta \Phi _g+R\right) \Delta \Theta _f^{\text{nnlo}}\right) \Delta \Theta _g+4 \Delta
\Theta _f \left(8 \Delta \Theta _f^{\text{nlo}}+\left(16 a_0+a_1\right) \Delta \Theta _f^{\text{nnlo}}\right) \Delta \Theta _g{}^2\Big)\Big\}\Big]\Bigg)\right/\\
&&\left(4 b_1 R \left(-4 b_1{}^2+\left(\Delta \Phi _f-\Delta \Phi _g\right){}^2+4 \Delta \Theta _f \Delta \Theta _g\right) \left(-b_1{}^2+\left(\Delta
\Phi _f-\Delta \Phi _g\right){}^2+4 \Delta \Theta _f \Delta \Theta _g\right)\right)
\end{eqnarray*}



\begin{eqnarray*}
R=\sqrt{\Delta \Phi _f{}^2-2 \Delta \Phi _f \Delta \Phi _g+\Delta \Phi _g{}^2+4 \Delta \Theta _f \Delta \Theta _g},.\hspace{7 cm}\text{(B.5)}
\end{eqnarray*}
\end{widetext}

%
%

\end{document}